\RequirePackage{fix-cm}
\documentclass[twocolumn,epjc3,showpacs,preprintnumbers,amsmath,amssymb]{svjour3}
\smartqed
\RequirePackage{graphicx}
\RequirePackage[numbers,sort&compress]{natbib}
\usepackage[colorlinks,linkcolor=blue,anchorcolor=blue,citecolor=green]{hyperref}
\usepackage{graphicx,amsmath,amsfonts,latexsym,amssymb,graphics,epsfig,subfigure,color,makeidx}
\usepackage{multirow,epstopdf}
\journalname{Eur. Phys. J. C}
\usepackage[title]{appendix}
\newcommand {\nn}{\nonumber}
\newcommand {\be}{\begin{equation}}
\newcommand {\ee}{\end{equation}}
\newcommand {\beq}{\begin{eqnarray}}
\newcommand {\eeq}{\end{eqnarray}}
\newcommand {\pa}{\partial}

\makeatletter

\begin{document}
\title{The effect of scalar hair on the charged black hole with the images from accretions disk}

\author{Tao-Tao Sui\thanksref{addr1}
   \and Zi-Liang Wang\thanksref{addr2}
   \and Wen-Di Guo\thanksref{e4,addr3}
}
\thankstext{e4}{e-mail:guowd@lzu.edu.cn, corresponding author}
\institute{College of Physics, Nanjing University of Aeronautics and Astronautics, Nanjing 211106, China  \label{addr1}
\and
Department of Physics, School of Science, Jiangsu University of Science and Technology, Zhenjiang, 212003, China \label{addr2}
\and
Institute of Theoretical Physics $\&$ Research Center of Gravitation, Lanzhou University, Lanzhou 730000, China \label{addr3}
}

\date{Received: date / Accepted: date}

\maketitle

\begin{abstract}
{In this paper, we investigate the optical properties of a charged black hole with scalar hair (CSH) within the context of four-dimensional Einstein-Maxwell-Dilaton gravity. To achieve this, we consider three distinct toy models of thin accretion disks. The presence of dilaton coupling allows us to express both the solutions of CSH and the Reissner-Nordström (RN) black hole in terms of their mass ($M$) and charge ($Q$). Our findings reveal differences in the effective potentials $V_{eff}$, photon sphere radii $r_{ph}$, and innermost stable circular orbit $r_{isco}$ between the CSH and RN black hole cases, which become increasingly pronounced as the charge parameter $Q$ increases. However, no noticeable distinctions are observed concerning the critical impact parameter $b_{ph}$. When the ratio of the photon ring band and the lensed ring band exceeds 0.1, it may suggest the presence of a charged black hole with scalar hair. Furthermore, our results underscore the significant influence of the charge parameter $Q$ on the brightness distributions of the direct, lensed ring, and photon ring for three standard emission functions. These findings emphasize the potential for distinguishing between CSH and RN black holes through an analysis of direct intensity and peak brightness in specific accretion disk models.}
\end{abstract}

\maketitle

\section{Introduction}
With the development of technology in recent years, more and more observations confirm the prediction of General Relativity (GR), which can result GR is a successful theory to describe the geometric properties of spacetime. Nonetheless, GR can not account a certain phenomena, such as the accelerated expansion of the universe \cite{SupernovaSearchTeam:1998fmf,SupernovaCosmologyProject:1998vns}, we need to acknowledge that our present comprehension of gravity may remain incomplete. Therefore, the  generally thought is that we can attempt to modify the GR, and give an explanation of this phenomena \cite{Clifton:2011jh}. 

Moreover, the no-hair theorem, one of the main characteristics in GR, states that the  stationary black hole can be only described by mass $(M)$, electric charge $(Q)$ and angular momentum $(a)$. For the modified gravity theory, the inclusion of modified terms introduce supplementary degrees of freedom and offer explanations for phenomena that GR cannot adequately address. Indeed, significant efforts have been made to modify GR by incorporating additional matter fields into the theories, which admits that the black holes need to be described by the other hair. Therefore, the hairy black holes have been widely constructed and analyzed, including the radially dependent scalar field \cite{Horndeski:1974wa,Rinaldi:2012vy,Cisterna:2014nua,Feng:2015oea,Sotiriou:2013qea, Miao:2016aol,Kuang:2016edj,Babichev:2016rlq,Benkel:2016rlz,Filios:2018xvy,Cisterna:2018hzf,Giusti:2021sku} and time-dependent scalar field \cite{Babichev:2013cya,Babichev:2017lmw,BenAchour:2018dap,Takahashi:2019oxz,Minamitsuji:2019shy,Arkani-Hamed:2003juy}. 

Recently, Bah and Heidmann introduced a modified gravity theory capable of describing a five-dimensional topological star/black hole model \cite{Bah:2020pdz,Bah:2020ogh}. This theory is founded on a five-dimensional Einstein-Maxwell framework. Within this model, the spacetime remains continuous and smooth in microstate geometries, while exhibiting similarity to a classical black hole in macrostate geometries. For more details, please refer to Refs. \cite{Bah:2022yji,Stotyn:2011tv,Heidmann:2022ehn}. 
Utilizing the Kaluza–Klein reduction, the five-dimensional Einstein–Maxwell theory can be reduced into an effective four-dimensional Einstein–Maxwell–Dilaton theory, which can obtain a static spherically symmetric charged black hole with scalar hair (CSH). In Ref. \cite{Guo:2022nto}, the authors extensively investigated the shadow of this charged black hole with scalar hair from the perspectives of four different types of observers. Guo et al. also examined the quasinormal modes (QNMs) of this four-dimensional charged black hole with scalar hair, and their findings indicate that the discrepancies in the frequencies of the fundamental QNMs between the CSH and the Reissner-Nordstr$\ddot{o}$m (RN) black holes are notably small for the angular number $l=2$ \cite{Guo:2022rms}.

On the other hand, the Event Horizon Telescope (EHT) has ushered in a novel observational perspective on the vicinity of black holes, providing us with images of the supermassive black holes in M87* \cite{EventHorizonTelescope:2019dse,EventHorizonTelescope:2019uob,EventHorizonTelescope:2019jan,EventHorizonTelescope:2019ths,EventHorizonTelescope:2019pgp,EventHorizonTelescope:2019ggy}, and in SgrA* \cite{EventHorizonTelescope:2022wkp,EventHorizonTelescope:2022apq,EventHorizonTelescope:2022wok,EventHorizonTelescope:2022exc,EventHorizonTelescope:2022urf,EventHorizonTelescope:2022xqj}. The black hole shadows and the observational data provided by the EHT can encode valuable spacetime information, which also can be applied in the ongoing exploration of black hole physics. These applications include parameter estimations for black holes \cite{Kumar:2018ple,Ghosh:2020spb,Afrin:2021imp,Ghosh:2022kit}, constraints on the size of extra dimensions \cite{Vagnozzi:2019apd,Banerjee:2019nnj,Tang:2022hsu}, as well as investigations into GR and alternative theories of gravity \cite{Mizuno:2018lxz,Psaltis:2018xkc,Stepanian:2021vvk,Younsi:2021dxe,KumarWalia:2022aop,Vagnozzi:2022moj,Meng:2022kjs,Kuang:2022ojj,Gussmann:2021mjj,Khodadi:2022pqh,Khodadi:2021gbc}. The success of these images and the potential future experimental opportunities in imaging have the capacity to usher in transformative research in gravitational phenomenology, which can serve as catalysts for our exploration of physics within the context of strong gravitational fields. Equally thrilling is the potential to observe ultra-compact objects, such as boson stars \cite{Schunck:2003kk}, soft hair around the horizon \cite{Hawking:2016msc}, gravastars \cite{Mazur:2004fk}, and fuzzballs \cite{Mathur:2005zp}. Investigating their gravitational signatures and observational distinctions promises to pave the way for novel gravity tests through direct observations.

Besides, the black hole images released by the EHT capture the optical characteristics of black holes, stemming from the bending of photons due to gravitational lensing. In particular, the luminous region surrounding the dark area in these images is emitted by the accretion material enveloping real astrophysical black holes. The distribution of this accretion matter significantly impacts the observable shape of the black hole, due to the variations in the distribution of such matter. Although replicating a realistic accretion disk in theoretical research remains a challenging task, the first image of a black hole with a thin accretion disk was analytically computed in \cite{Luminet:1979nyg}. This analysis revealed the presence of primary and secondary images that appear outside the black hole shadow. Subsequently, in \cite{Bambi:2013nla}, it was pointed out that it is relatively straightforward to distinguish a Schwarzschild black hole from a static wormhole based on the characteristics of their shadow images. In Ref. \cite{Gralla:2019xty}, the authors focused the investigations into Schwarzschild black hole with both thin and thick accretion disks, the result shows that the lensed ring, along with the photon ring, contributes additional observable flux to the image. Another type of accretion, spherical accretion, has been employed to analyze the images of Schwarzschild black hole in Refs. \cite{Falcke:1999pj,Narayan:2019imo}.  These studies have revealed that the black hole shadows remain a robust feature, and its size and shape are primarily influenced by the spacetime geometry rather than the specific details of the accretion. Presently, research on photon rings and the observational aspects of black holes have garnered substantial attention, unveiling images of static black holes surrounded by various accretion scenarios that extend beyond GR \cite{Zeng:2020dco,Zeng:2020vsj,Peng:2020wun,Saurabh:2020zqg,Qin:2020xzu,Gan:2021pwu,Okyay:2021nnh,Li:2021ypw,Li:2021riw,Guo:2021bhr,Hu:2022lek,Guo:2021bwr,Wen:2022hkv,Chakhchi:2022fls,Hou:2022eev,Kuang:2022xjp,Uniyal:2022vdu,Uniyal:2023inx,Wang:2023vcv}.

{Return back to the four-dimensional Einstein–Maxwell-Dilaton model, the inclusion of dilaton coupling leads to a modification in the expression of the solution for charged black hole, without introducing any scalar charges. This observation suggests that the black hole solutions for both the CSH and RN cases can be described using the same parameters, namely, mass ($M$) and electric charge ($Q$). Nevertheless, the distinctions between the two solutions become evident in the appearance of the images of black holes. This revelation serves as an inspiration for the potential to differentiate between CSH and RN black holes based on their respective images. The principal aim of this study is to elucidate the influence of scalar hair on static black hole within the CSH scenario, and to discern the disparities between CSH and RN black holes by conducting an analysis of their optical characteristics when surrounded by three kinds of thin accretion disks.}

The paper is structured as follows: In Section \ref{themodel}, we provide an overview of charged black hole with scalar hair. In Section \ref{sec-Trajectories}, we delve into the calculation of photon trajectories of the CSH black hole. This analysis is carried out using Lagrangian formalism and a ray-tracing methodology. In Section \ref{thin}, we investigate the total observed intensities and explore the possible sources contributing to it for three standard disk accretion functions. Subsequently, we compare the optical appearances of the CSH black hole to that of the RN black hole. Finally, we present our concluding remarks and summarize the key findings in Section \ref{conclusion}.

\section{The review of charged black hole with scalar hair}\label{themodel}
First of all, we will provide a concise overview of the charged black hole with scalar hair. We shall consider a five-dimensional Einstein-Maxwell theory, wherein the complete action can be expressed in \cite{Bah:2020pdz} as
\begin{equation}
S_5=\int d^5x\sqrt{-\hat{g}}\left(\frac{1}{2\kappa_5^2}\hat{R}-\frac{1}{2}\hat{F}^{MN}\hat{F}_{MN}\right),
\end{equation}
where $\kappa_{5}$ can be considered as the five-dimensional gravitational constant and $\hat{F}^{MN}$ describes electric three-form field strengths. Here, we ignored the magnetic term. For simplify, the capital Latin letters $M, N...$ denote the five-dimensional coordinates, and the Greek letters $\mu, \nu...$ denote the four-dimensional coordinates. The corresponding spherical symmetric metric for the five-dimensional Einstein-Maxwell theory can be expressed as \cite{Bah:2020pdz,Stotyn:2011tv}
\begin{eqnarray}
ds^2&=&-f_S(r)dt^2+f_B(r)dy^2+\frac{1}{f_S(r)f_B(r)}dr^2\nn \\
&+&r^2(d\theta^2+\sin^2\theta d\psi^2), \label{metric_five}
\end{eqnarray}
where 
\begin{eqnarray}
f_{S}(r)=1-\frac{r_S}{r},~f_{B}(r)=1-\frac{r_B}{r},
\end{eqnarray}
and the five-dimensional vector field strength can be expressed as $\hat{F}=\frac{\hat{Q}}{r^2}dr\wedge dt\wedge dy$. Besides, the five-dimensional spacetime can be considered as a $S^1$ fibration over a four-dimensional Minkowski, $R^4\times S^{1}$, where the extra dimension is denoted by ``y'' with the radius $R_y$.

From the expression of the metric, we can see that the solutions exhibit a curvature singularity at $r=0$, along with two coordinate singularities located at $r=r_B$ and $r=r_S$. Ibrahima Bah et al. found a smooth bubble locates at $r=r_B$ \cite{Bah:2020pdz,Bah:2020ogh}. They also discussed the cases $r_S\geq r_{B}$ and $r_S<r_{B}$, which correspond to a black string and a topological star, respectively.

By the Kaluza-Klein reduction, which can be handled as integrating the extra dimension $y$, the five dimensional Einstein-Maxwell theory can be reduced into a four-dimensional Einstein-Maxwell-Dilaton theory as
\begin{eqnarray}\label{action_four}
S_4=\int d^4x\sqrt{-{g}}\Big(\frac{1}{2\kappa_4^2}R_4-\frac{3}{\kappa_4^2}\pa_{\mu}\Phi\pa^{\mu}\Phi -\frac{e^{2\Phi}}{2e^2} F_{\mu\nu}F^{\mu\nu}\Big),
\end{eqnarray}
where $\kappa_{4}^{2}=e^2\kappa_{5}^{2}$ and $e^2=\frac{1}{2\pi R_y}$. The corresponding scalar field can be defined as $\Phi=-\frac{1}{4}\ln(f_B)$, the four-dimensional field strength of vector field can be solved as $F=\frac{\hat{Q}}{r^2}dr\wedge dt$, 
and the four-dimensional metric can be rewritten as
\begin{eqnarray}
ds_4^2=f_B^{\frac{1}{2}}\Big(-f_Sdt^2+\frac{dr^2}{f_Bf_S}+r^2d\theta^2+r^2\sin^2\theta d\psi^2\Big)\label{metric_four}.
\end{eqnarray}
From the four-dimensional perspective, the conserved quantities, the ADM mass $M$ and the electric charge $Q$ can be given as 
\begin{eqnarray}
M=\frac{2\pi}{\kappa_4^2}(2r_S+r_B),~Q^2=\frac{\hat{Q}^2}{e^2}=\frac{3r_S r_B}{2\kappa_4^2}.
\end{eqnarray}
For convenience, we can express two pairs $(r_S, r_B)$ for given mass and charge parameters $(M, Q)$:
\begin{eqnarray}
r_S^{(1)}&=&\frac{\kappa_4^2}{8\pi}(M+M_{\delta}),~r_B^{(1)}=\frac{\kappa_4^2}{4\pi}(M-M_{\delta}),\label{case1}\\
r_S^{(2)}&=&\frac{\kappa_4^2}{8\pi}(M-M_{\delta}),~r_B^{(2)}=\frac{\kappa_4^2}{4\pi}(M+M_{\delta}),\\
M_{\delta}&=&\sqrt{M^2-(8\pi Q)^2/(3\kappa_4^2)}.\nn
\end{eqnarray}
{While, the corresponding event horizons of the Reissner-Norstr$\ddot{o}$m black hole are given as
\begin{eqnarray}
r_{\pm}^{RN}=\frac{\kappa_4^2}{8\pi}(M\pm \sqrt{M^2-(4\sqrt{2}\pi Q)^2/\kappa_4^2}).
\end{eqnarray} 
Here, we should note that the RN black hole is not the solution of  Einstein-Maxwell-Dilaton theory as Eq. \eqref{action_four}}. To facilitate our calculations, we can set the electric charge range of the RN black hole as $Q\in[0,M]$. On the other hand, the corresponding electric charge range of the CSH black hole is specified as $Q\in[0,\sqrt{{}3/2}M]$. Besides, the authors studied shadow of the CSH black bole, revealing that the radial component of geodesic equation for the photon is influenced by $f_S(r)$ in Ref. \cite{Guo:2022nto}. Consequently, we will exclusively focus on the scenario denoted by \eqref{case1} with the electric charge $Q\in[0, 2/\sqrt{3}M]$, and this range remains greater than that of the RN black hole's electric charge. {In fact, when we observed a charged black hole, we can not make sure which gravitational theory solution the charged black hole is. So, we need to distinguish between observations of black holes with the same charge, which means that the same charge will give different observed effects on the different charged black hole modes. Hence,} in the next section, we will discern the impact of scalar hair on the image of the static black hole by investigating the observable properties of CSH and RN black holes when subjected to various accretion processes.

\section{Trajectories of photons around charged black hole with scalar hair}\label{sec-Trajectories}

\begin{figure}[htbp]
\centering 
{\includegraphics[width=0.33\textwidth]{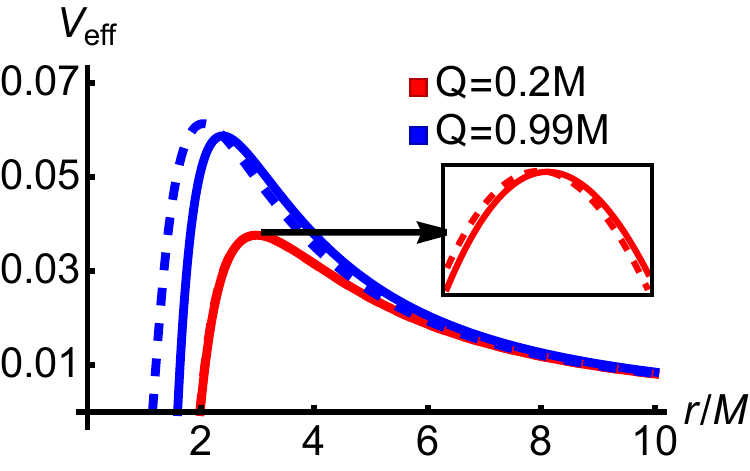}}
\caption{The behaviors of effective potentials $V_{\text{eff}}$ for the CSH black hole (the solid lines) and RN black hole (the dashed lines).}\label{figveff}
\end{figure}

In this section, we will explore the behavior of light rays near the CSH black hole, specifically focusing on null geodesics. From a geometric perspective, the paths of photons correspond to null geodesics of the spacetime. Consequently, we must address the null geodesic equations. However, it's important to note that the geodesic equations governing the motion of photons involve four coupled second-order differential equations, making direct solutions challenging. An alternative approach involves encapsulating the geodesic equation of photons within the framework of the Euler–Lagrange equation:
\begin{equation}\label{eq-ELeq}
\frac{d}{d\lambda}\left(\frac{\partial \mathcal{L}}{\partial \dot{x}^\mu}\right)=\frac{\partial \mathcal{L}}{\partial x^\mu},
\end{equation}
where $\lambda$ is the affine parameter and $\dot{x}^\mu=d{x}^\mu/d\lambda$ represents the four-velocity of the photon. The photons in the spacetime take the Lagrangian
\begin{eqnarray}
\mathcal{L}&=&\frac{1}{2}g_{\mu\nu}\dot{x}^\mu\dot{x}^\nu\nn\\
&=&\frac{1}{2}f_B^{\frac{1}{2}}\Big(-f_S\dot{t}^2+\frac{\dot{r}^2}{f_Bf_S}+r^2(\dot{\theta}^2+\sin^2\theta \dot{\psi}^2)\Big).\label{lag}
\end{eqnarray}
Because of the spherical symmetry of the spacetime, we can focus on the photons moving on the equatorial plane, i.e., $\theta=\pi/2$ and $\dot{\theta}=0$. For this static black hole, there are two Killing vectors $\partial_{t}$ and $\partial_{\psi}$, which lead to the conserved total energy $E$ and angular momentum $L_z$,
\begin{equation}
E\equiv -\frac{\partial \mathcal{L}}{\partial \dot{t}}=\sqrt{f_{B}}f_{S}\dot{t},   \qquad  L_z\equiv \frac{\partial \mathcal{L}}{\partial \dot{\psi}}=\sqrt{f_{B}}r^2\dot{\psi}.\label{conserved}
\end{equation}
By redefining the affine parameter $\bar{\lambda}$ as $\lambda/L_z$ and the impact parameter $b\equiv L_z/E$, and recalling $\mathcal{L}=0$ for photons, we can reduce Eq. \eqref{eq-ELeq} into three first-order differential equations, which are the time, azimuthal and radial components of the four-velocity 
\begin{align}
&\dot{t}=\frac{1}{b \sqrt{f_{B}}f_{S}},\label{eq1}\\
&\dot{\psi}=\pm\frac{1}{\sqrt{f_{B}}r^2},\label{eq2}\\
&\dot{r}^2=\frac{1}{b^2}-\frac{f_{S}}{r^2}=\frac{1}{b^2}-V_{\text{eff}}(r),\label{eq3}
\end{align}
where the sign $``+"$ and $``-"$ of azimuthal component $\dot{\psi}$ denote that the light ray traveling in the counterclockwise and clockwise along azimuthal direction. The effective potential $V_{\text{eff}}(r)$ is illustrated in Fig. \ref{figveff}. This plot reveals that the effective potential vanishes initially at the event horizon. Subsequently, as $r$ increases, it attains a maximum value before gradually decreasing monotonically. Besides, as the charge parameter $Q$ increases, the distinction between the effective potentials for the CSH and RN black holes becomes increasingly evident, which seems to indicate that motion of photons becomes distinguishable in the two types of black holes for large charge parameter $Q$.

\begin{figure}[htbp!]
\begin{center}
\subfigure[]{\label{rph}
\includegraphics[width=0.23\textwidth]{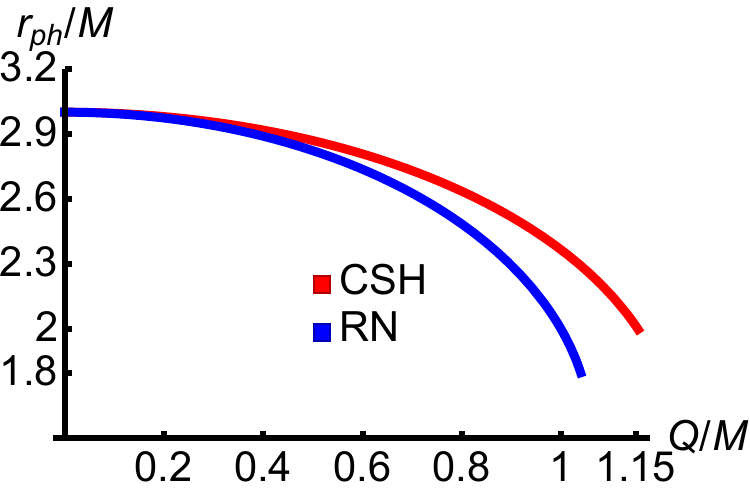}}
\subfigure[]{\label{bph}
\includegraphics[width=0.23\textwidth]{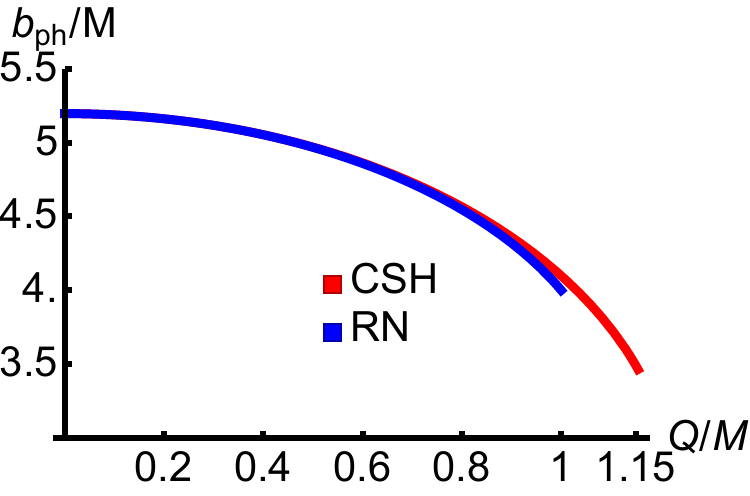}}
\caption{The radius of photon sphere $r_{ph}$ and the critical impact parameter $b_{ph}$ for the CSH black hole (the red lines) and RN black hole (the blue lines).}\label{rphbph}
\end{center}
\end{figure}

Prior to delve into the comprehensive photon trajectories, it is crucial to conduct an analysis of the circular orbit or photon sphere. This orbit is characterized by the conditions $\dot{r}=0$ and $\ddot{r}=0$, as indicated by Eq. \eqref{eq3}. Using translating the conditions of the photon sphere, we can derive
\begin{equation}
V_{\text{eff}}(r_{ph})=\frac{1}{b_{ph}^2}, \qquad V_{\text{eff}}'(r_{ph})=0. \label{formulabrph}
\end{equation}
where the prime notation signifies differentiation with respect to $r$. Based on the above equations, the radius of photon sphere $r_{ph}$ and the critical impact parameter $b_{ph}$ can be obtained for different values of charge parameter $Q$, which are shown in Fig. \ref{rphbph}. From Fig. \ref{rphbph}, we can see that both the radius of photon sphere $r_{ph}$ and the critical impact parameter $b_{ph}$ decrease with the charge parameter $Q$, a trend observed in both CSH and RN black hole cases. It's worth highlighting that the rate of decline for the radius of photon sphere $r_{ph}$with respect to the charge parameter $Q$ is less pronounced in the context of the CSH black hole, as opposed to the RN black hole. While, for the critical impact parameter $b_{ph}$, the CSH and RN black holes do not exhibit a significant difference, in the electric charge range $Q\in[0,M]$. 

Besides, it's essential to recognize that there are three distinct scenarios for different values of the impact parameter $b$: i) $b>b_{ph}$, the photon will scatter 
into infinity after traversing the turning point; ii) $b=b_{ph}$, the photon does not escape to infinity but instead travels around the black hole along a circular orbit with a radius of $r_{ph}$, effectively forming the boundary of the black hole's shadow; iii) $b<b_{ph}$, the photon inexorably approaches to the event horizon, getting closer and closer until it ultimately falls into the black hole.

Furthermore, by combining Eq. (\ref{eq2}) and Eq. (\ref{eq3}), we can derive an equation of motion that accurately portrays the trajectory of the light ray
\begin{equation}
\frac{dr}{d\psi}=\pm \sqrt{f_{B}(r)}r^2\sqrt{\frac{1}{b^2}-V_{\text{eff}}(r)}.\label{eqphoton}
\end{equation}
The total change in the azimuthal angle $\psi$ depends on the impact parameter $b$ as described in Eq. \eqref{eqphoton}.  To distinguish between the photon ring and the lensing ring near the black hole, Wald et al. introduced the concept of the orbit number, denoted as $n=\psi/(2\pi)$, which can quantify the total count of loops executed by the photon around the black hole \cite{Gralla:2019xty}. They categorized photon orbit trajectories into three primary cases based on the orbit number $n$. More specifically, the first category is characterized as direct emission with $n<3/4$, denoting that the light ray intersects the accretion disk solely once. The second category, encompassing $3/4<n<5/4$, corresponds to the lensed ring emission, where the light ray intersects the accretion disk twice. The third classification, referred to as photon ring emission, applies to light rays with $n>5/4$, which cross the accretion disk a minimum of three times. To gain a more comprehensive understanding of these three trajectory classes, schematic diagrams can be referred to in \cite{Wielgus:2021peu, Bisnovatyi-Kogan:2022ujt, Hu:2022lek}.

\begin{figure}[htbp!]
\begin{center}
\subfigure[$Q=0.6M$]{\label{trajectory2}
\includegraphics[width=0.22\textwidth]{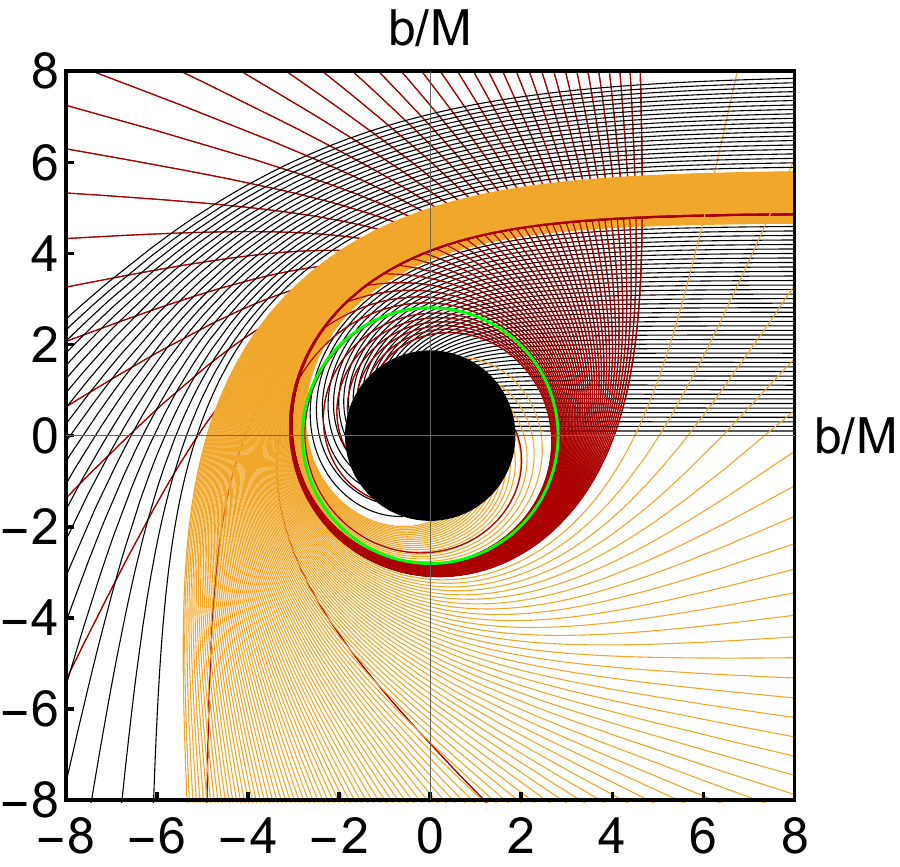}}
\subfigure[$Q=1.15M$]{\label{trajectory9}
\includegraphics[width=0.22\textwidth]{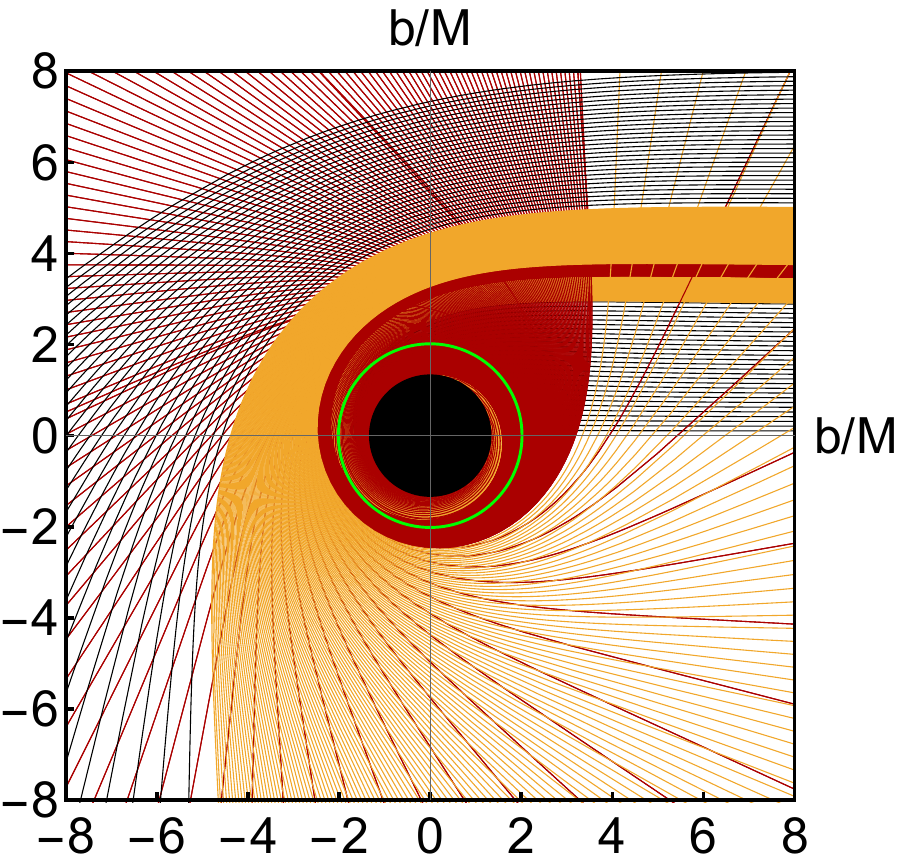}}
\caption{The trajectories of photons, described with the impact parameter b, are shown in $(r,\psi)$ polar coordinates of the CSH black hole. The black, gold, and red curves correspond to the direct $(n<3/4)$, lensed ring $(3/4<n<5/4)$, and photon ring $(n>5/4)$, respectively.}\label{trajectory}
\end{center}
\end{figure}

Figure \ref{trajectory} presents the trajectories of photons with different impact parameters $b/M$ in the polar coordinates $(r,\psi)$. In this depiction, we consider an observer locates at a considerable distance to the right of the plot. The black curves represent direct emission, the gold curves correspond to lensed ring emission, the red curves indicate photon ring emission, and the green circular ring describes the photon sphere. From Fig. \ref{trajectory}, it's evident that the charge parameter $Q$ has a notable impact on the size and distribution of the direct, lensed, and photon ring bands. For example, as the charge parameter $Q$ increases, the photon ring band tends to become larger.

\begin{figure}[htbp!]
\begin{center}
\subfigure[$Q=0.6M$]{\label{orbitn2}
\includegraphics[width=0.23\textwidth]{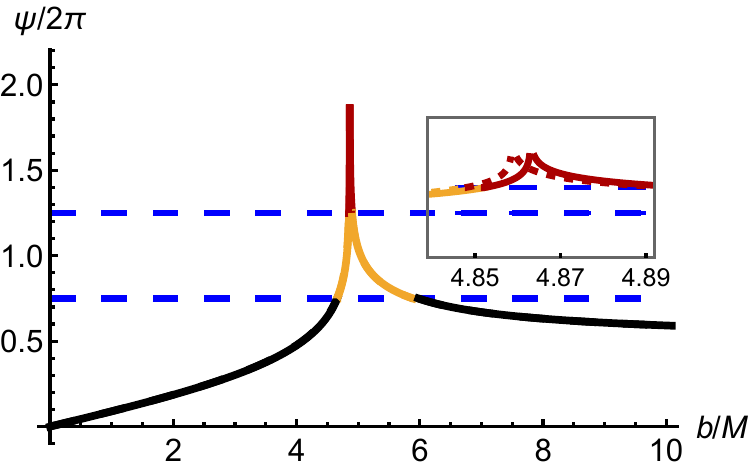}}
\subfigure[$Q=0.99M$]{\label{orbitn9}
\includegraphics[width=0.23\textwidth]{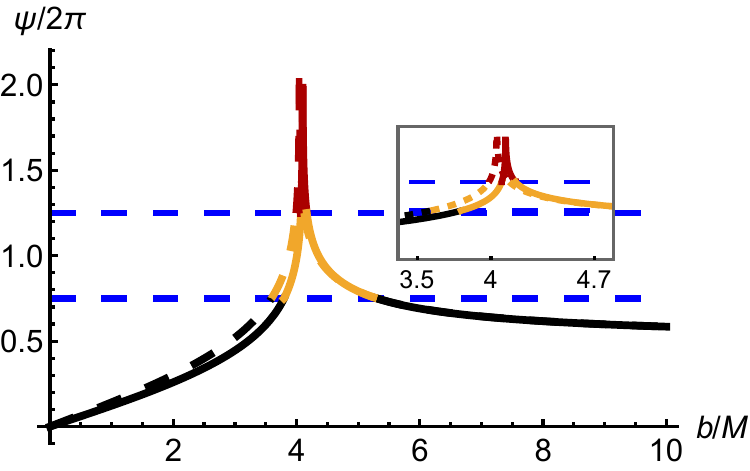}}
\caption{The orbit numbers $n$ as the functions of impact parameter b with different charge parameter $Q$ for CSH (the solid line) and RN (the dashed line) black hole cases. The black, gold, and red curves correspond to the direct $(n<3/4)$, lensed ring $(3/4<n<5/4)$, and photon ring $(n>5/4)$ emissions respectively.}\label{orbitn}
\end{center}
\end{figure}

\begin{figure}[htbp!]
\begin{center}
\subfigure[$n>3/4$]{\label{n34}
\includegraphics[width=0.23\textwidth]{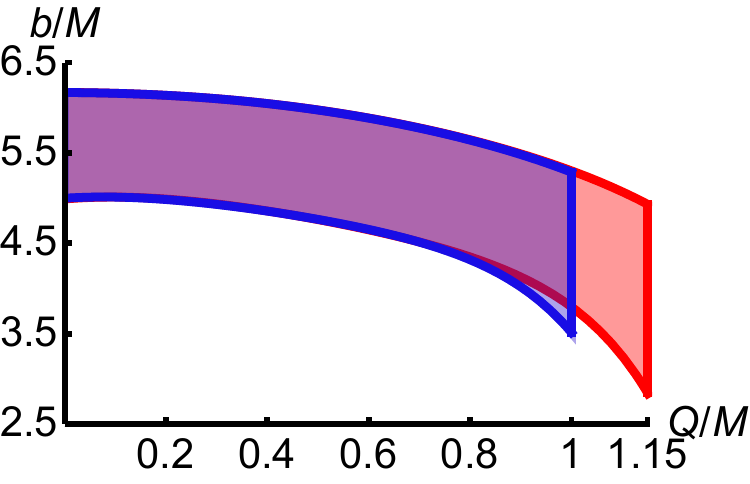}}
\subfigure[$n>5/4$]{\label{n54}
\includegraphics[width=0.23\textwidth]{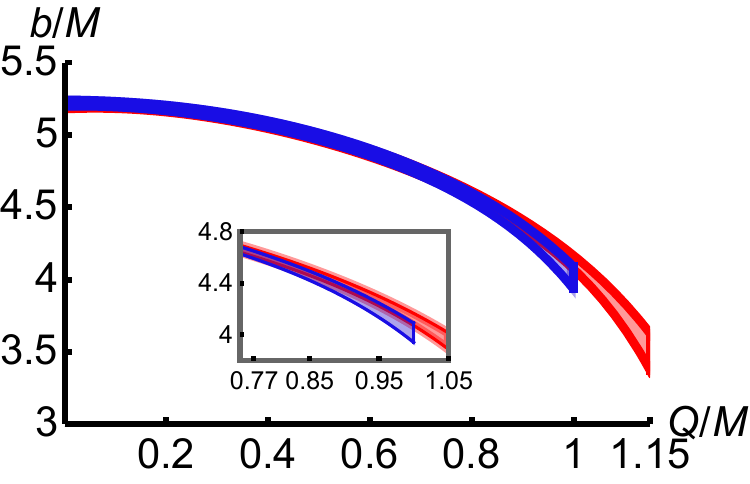}}
\subfigure[$\Gamma$]{\label{ratiopl}
\includegraphics[width=0.23\textwidth]{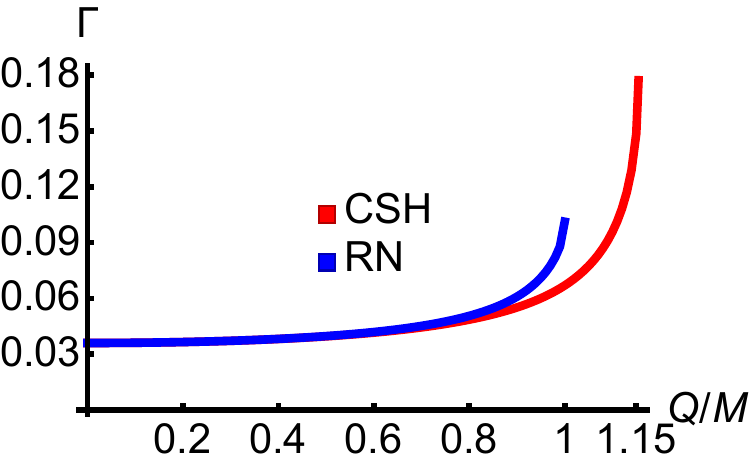}}
\caption{The range of the impact parameter b corresponding to $(n>3/4)$, and $(n>5/4)$ and the ratios $\Gamma$ for the CSH black hole case and the RN black hole case for different charge parameter $Q$. The red regions denote the CSH black hole case, and the blue parts represent the RN black hole case.}
\end{center}
\end{figure}

Besides, we provide the orbit photon sphere number $n$ as a function of the impact parameter $b$ for charge parameter $(Q=0.6M, 0.99M)$ in the cases of the CSH and RN black holes. Figure \ref{orbitn} illustrates that as the charge parameter $Q$ increases, it becomes feasible to distinguish between the two black hole scenarios based on the orbit number $n$. Additionally, we show the range of the impact parameter $b$ corresponding to $(n>3/4)$ and $(n>5/4)$, which are the shaded areas in Fig \ref{n34} and \ref{n54}. They also illustrate that the differentiation between the two black hole cases becomes apparent as the charge parameter $Q$ approaches $M$. {To characterize the effect of charge parameter $Q$ on the orbit number $n$, we define a parameter $\Gamma$ to describe the ratio between the photon ring and the lensed ring bands for various charge parameter $Q$ as
\begin{equation}\label{gamma}
\Gamma\equiv\frac{\text{range of $b$ for $(n>5/4)$} }{\text{range of $b$ for $(5/4>n>3/4)$}},
\end{equation}
which can considered as a simply theoretical tool to differentiate the influence of charge parameter $Q$ on the charged black holes, and the notable insights can be derived from Fig. \ref{ratiopl}}. It is apparent that the ratios $\Gamma$ for the CSH  and the RN black hole cases experience an increase as the charge parameter $Q$ grows. For larger values of $Q$, the photon ring band becomes more prominent, consistent with the trends in Fig. \ref{trajectory}. Additionally, it is noteworthy to highlight that the increase of $\Gamma$ for both the CSH and RN black hole scenarios is practically identical within the range of the charge parameter $Q\in[0,0.8M]$. However, when the charge parameter extends to $Q\in[0.8M,M]$, the augmentation of $\Gamma$ becomes more pronounced in the RN black hole case as compared to the growth pattern observed in the CSH black hole scenario. In simpler terms, higher values of the charge parameter $Q$ result to a clearer distinction between the two black hole scenarios based on the $\Gamma$ ratio.

\section{Rings and images of hairy black hole illuminated by thin disk accretions}\label{thin}

In the previous section, we investigated the impact of the charge parameter $Q$ on photon trajectories and categorized the photons into three distinct classes based on their orbit numbers $n$. In the following section, we will focus on the exploration of imagery involving the charged black hole with scalar hair. The investigation centers around scenarios where the black hole is illuminated by optically and geometrically thin accretion disks. These disks are situated at rest on the equatorial plane surrounding the black hole and are observed face-on. In this context, as light rays traverse through the accretion disk, they interact with it, extracting energy in the process. It's noteworthy that various types of light rays make distinct contributions to the observed light intensity.

We can assert that the thin accretion disk emits intensity $I_e(r)$ with the frequency $\nu_e$. According to Liouville’s theorem, the specific intensity $I_{o}(r)$ received by the observer with an emission frequency $\nu_o$ can be expressed as \cite{Gralla:2019xty}
\begin{equation}
I_{o}(r, \nu_o)=g_s^3 I_{e}(r,\nu_e),
\end{equation}
where the factor $g_s= \nu_o/\nu_e=\sqrt{g_{tt}}$ represents the red shift factor. The total observed intensity, denoted as $I_{obs}(r)$, can be obtained by integrating over all observed frequencies of $I_{o}(r, \nu_o)$, which can be expressed as:
\begin{eqnarray}
I_{obs}(r)&=&\int I_{o}(r,\nu_o)d\nu_o=\int g_s^4 I_{e}(r,\nu_e) d\nu_e\\
 &=&(g_{tt}^2)\int I_{e}(r,\nu_e) d\nu_e=(g_{tt}^2) I_{em}(r),
\end{eqnarray}
where $I_{em}(r)=\int I_{e}(r,\nu_e) d\nu_e$. We should note that everytime the photon passes through the disk, it will extract energy from the accretion disk and contribute brightness to the observer. As we discussed in the previous subsection, a light ray can pick up brightness from the accretion disk once, twice, or even more times which is contingent upon its impact parameter $b$. Hence, the total observed intensity is the cumulative sum of intensities resulting from each intersection \cite{Gralla:2019xty}, yielding: 
\begin{equation}
I_{obs}(b)=\sum_{m}g_{tt}^2I_{em}(r)|_{r=r_m (b)}, \label{eqtransfer}
\end{equation} 
where $r_m(b)$ serves to characterize the transformation from the light ray's impact parameter $b$ to the radial coordinate of the $m$-th intersection with the accretion disk, which is also named as transfer function. Moreover, the slope of the transfer function $dr/db$ can describe the demagnification factor at impact parameter $b$.

\begin{figure}[htbp!]
\centering
\subfigure[$Q=0.6M$]
{\includegraphics[width=0.232\textwidth]{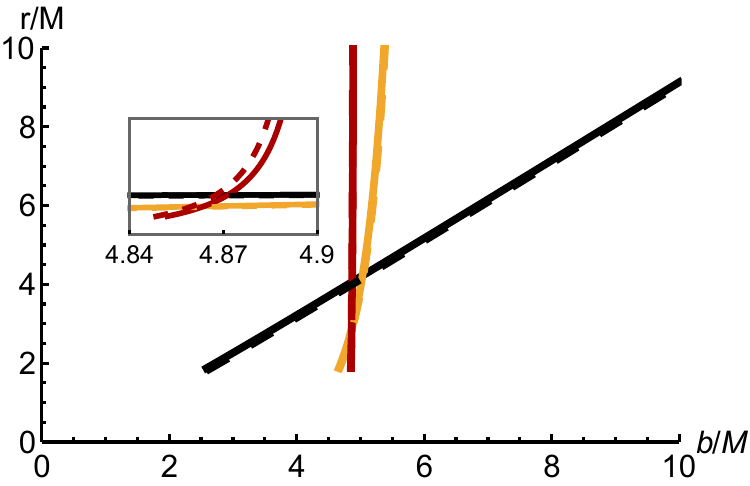}}
\subfigure[$Q=0.99M$]
{\includegraphics[width=0.232\textwidth]{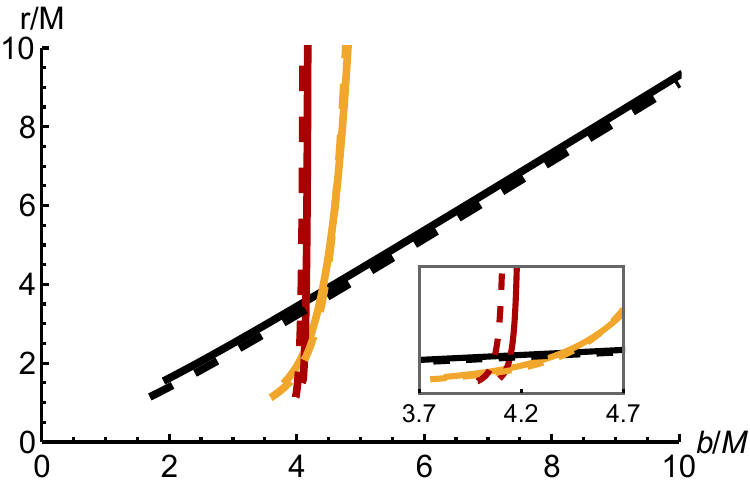}}
\caption{The first three transfer functions in CSH (the solid line) and RN (the dashed line) black holes for different charge parameter $Q$. They represent the radial coordinate of the first (black), second (gold), and third (red) intersections with the emission disk.}\label{figtransfer}
\end{figure}

Figure \ref{figtransfer} shows the first three transfer functions $r_m (b)$ with different charge parameter $Q$ for the two black hole cases. {The first transfer function (the black curves) labels that the photon only interacts with the disk one time.} Its slope remains nearly a constant, which approximats to 1, indicating that the direct image profile primarily mirrors the red shifted source profile. {For the second transfer function, colored with gold curves, is associated with the lensed ring emissions.} Due to its steep slope, an observer will perceive a significantly demagnified image of the accretion disk. Finally, as for the third transfer function, represented by the red curves, it corresponds to the photon ring emission. It is apparent that this image is subject to extreme demagnification. Consequently, it makes a minimal contribution to the total flux and is scarcely visible. Besides, the charge parameter $Q$ has small influence on the slope of the transfer function $r_m(b)$. However, owing to the same impact of the charge parameter $Q$ has on the critical impact factor $b_{ph}$ for the two black hole scenarios with $Q\in[0,M]$, the difference in the transfer function $r_m(b)$ between the two black hole cases is not substantial.

In order to distinguish the difference between the two black hole cases which can base on the specific intensity distribution around the black hole on the equatorial plane, we turn our attention to the optically and geometrically thin disk models with different emission functions. It's important to note that these models are particularly suitable for evaluating accretion processes where flow velocities are sub-Eddington and opacity is high, while they may not account for scenarios involving high accretion velocities and mass \cite{Page:1974he}. For example, around supermassive black holes, the accretion disk might effectively transit into an apparently thin but structurally thick configuration \cite{Guerrero:2021ues}. In this analysis, the primary source of observed specific light intensity is a thin disk, and its luminosity exclusively relies on the radial coordinate, denoted as $r$. As a result, we will consider the following three toy models \cite{Wang:2022yvi,Yang:2022btw}, which are employed to approximate real-world scenarios involving thin matter distributions.

\textbf{Model a:} The emission function of the accretion disk  starts from the innermost stable circular orbit (ISCO). This function has a square decay behavior,  which can be written as 
\begin{align}
\quad I^a_{em}(r)&:=
\begin{cases}
I_0\left[\frac{1}{r-(r_{ isco}-1)}\right]^2, &\hspace{0.5cm}  r>r_{isco}\\
0,&\hspace{0.5cm} r \leqslant r_{ isco}
\end{cases}\label{diskprofile1}
\end{align}
where $I_0$ is the maximum intensity and the $r_{isco}$ is the ISCO of the black hole. 

\textbf{Model b:} The emission function has a cubic decay behavior which starts from the photon sphere $r_{ ph}$
\begin{align}
\quad I^b_{ em}(r)&:=
\begin{cases}
I_0\left[\frac{1}{r-(r_{ ph}-1)}\right]^3, &\hspace{0.5cm}  r>r_{ ph}\\
0,&\hspace{0.5cm} r\leqslant r_{\rm ph}
 \end{cases}\label{diskprofile2}
\end{align}

\textbf{Model c:} The emission function is moderate decay, which starts from the event horizon $r_h$ 
\begin{align}
\quad I^c_{em}(r)&:=
\begin{cases}
I_0\frac{\frac{\pi}{2}-\arctan[r-(r_{ isco}-1)]}{\frac{\pi}{2}-\arctan[r_{ h}-(r_{ isco}-1)]}, &\quad r>r_{ h}\\
0,&\quad r\leqslant r_{ h}
\end{cases}.\label{diskprofile3}
\end{align}

\begin{figure}[htbp!]
\centering
\subfigure[$Q=0.2M$]{\label{modela}
\includegraphics[width=0.232\textwidth]{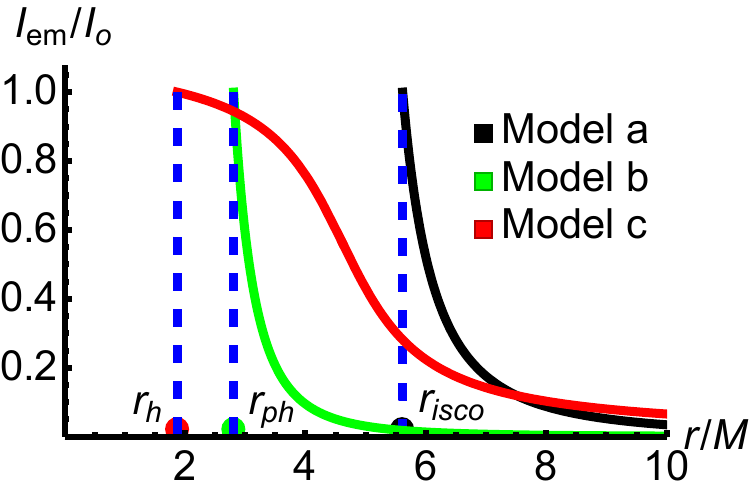}}
\subfigure[$r_{isco}$]{\label{risco}
\includegraphics[width=0.232\textwidth]{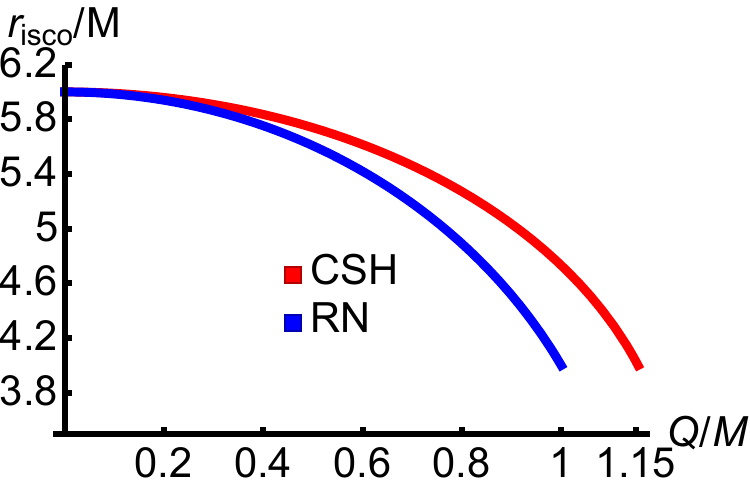}}
\caption{The emission function of the accretion disk for the three models with the charge parameter $Q=0.2M$ and the innermost stable circular orbit $r_{isco}$ for CSH (the red line) and RN (the blue line) black hole cases.}\label{emissionrisco}
\end{figure}

Figure \ref{modela} shows the sketches of the emission functions of the accretion disk for the three models. Figure \ref{risco} depicts the behavior of the $r_{isco}$ for CSH and RN black holes, it becomes evident that as the charge parameter $Q$ increases, the difference in $r_{isco}$ between the CSH and RN black hole cases becomes more pronounced. Besides, the innermost stable circular orbit of RN black hole case is consistently closer to the event horizon than in the CSH black hole case.

\begin{figure}[htbp!]
\begin{center}
\subfigure[]{\label{three ointensities}
\includegraphics[width=0.23\textwidth]{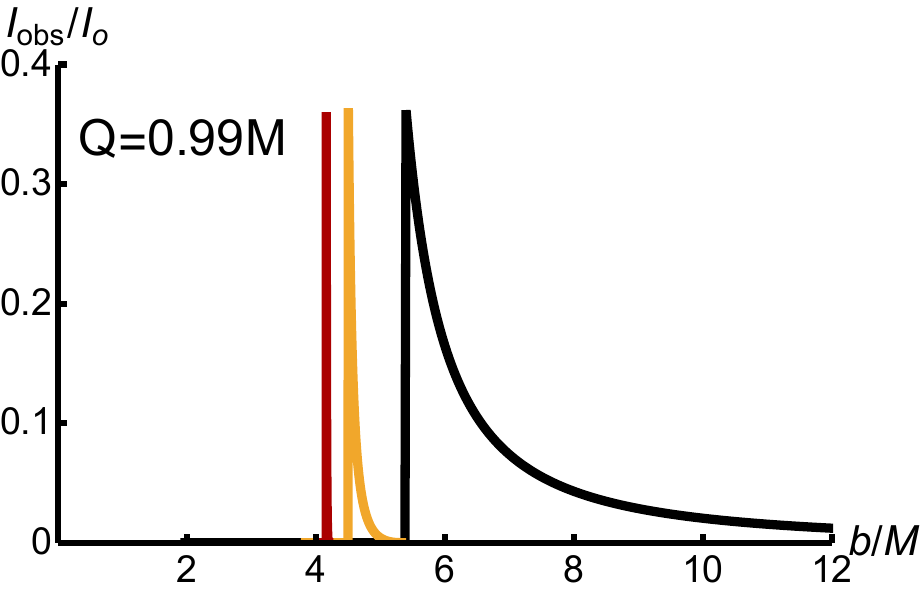}}
\subfigure[]{\label{total ointensities}
\includegraphics[width=0.23\textwidth]{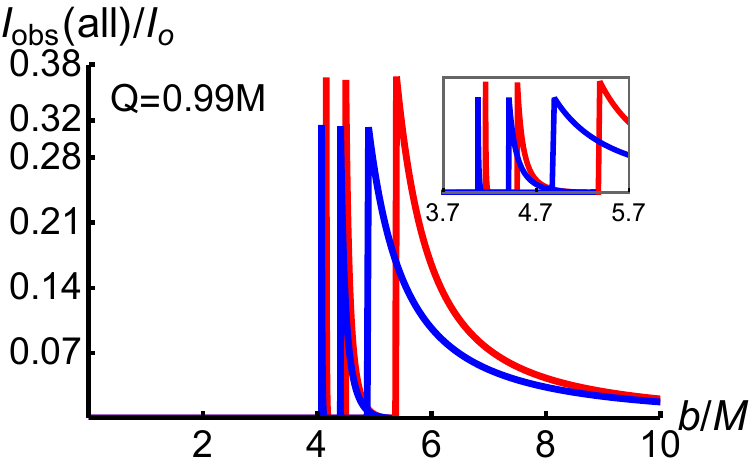}}
\subfigure[]{\label{disksh1}
\includegraphics[width=0.2\textwidth]{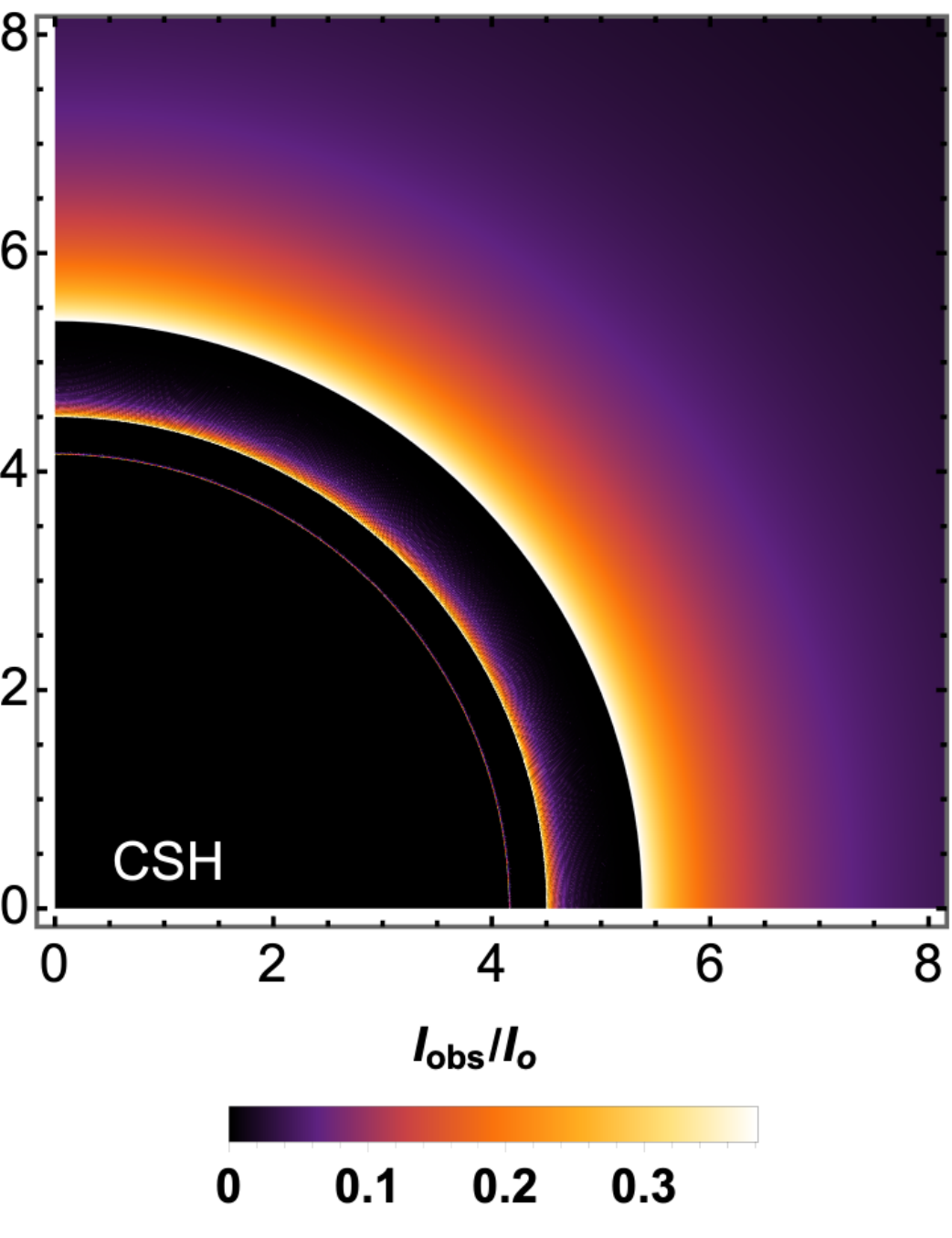}\hspace{1cm}
\includegraphics[width=0.2\textwidth]{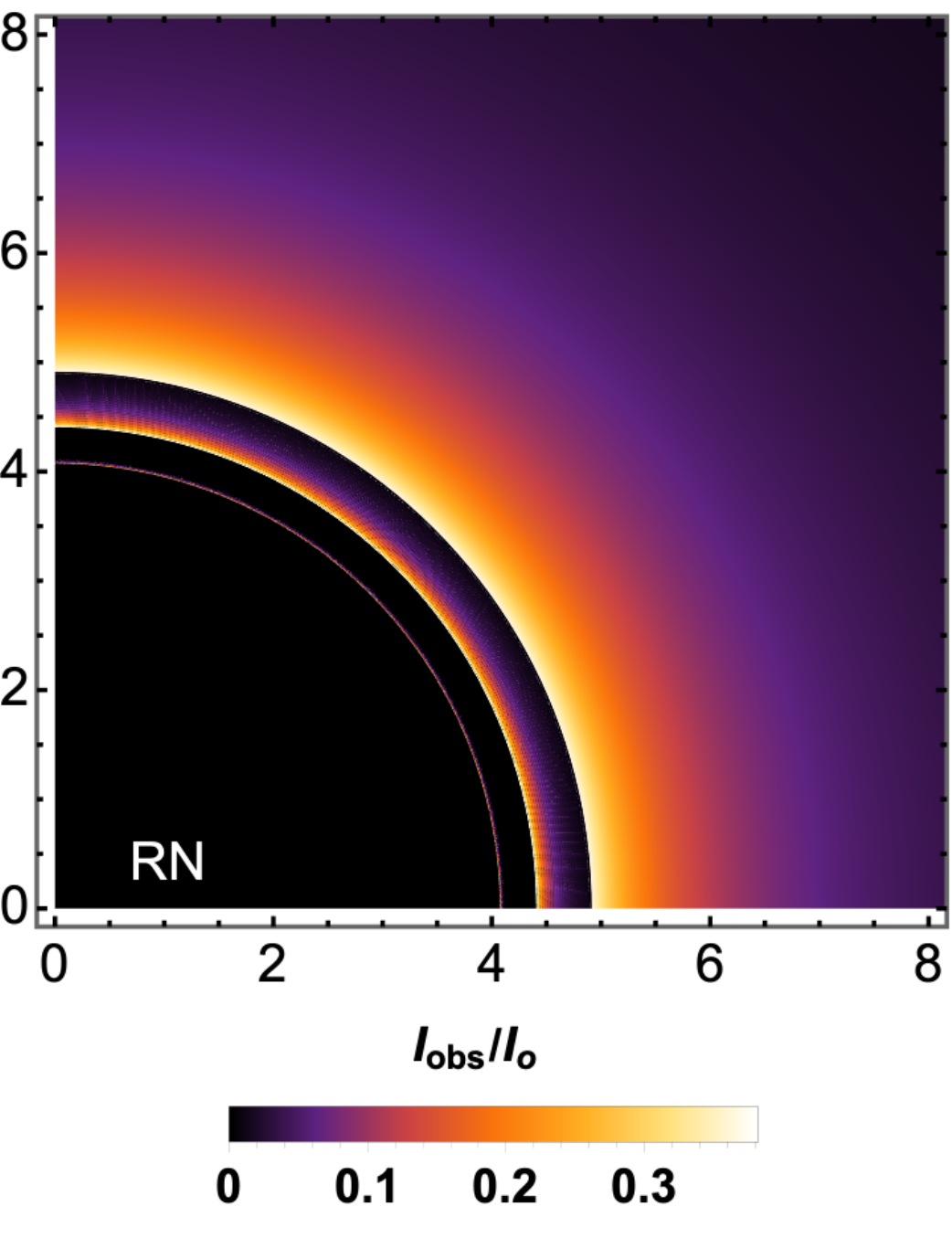}}
\caption{Observational appearances of the thin disk with \textbf{Model a} for the CSH and RN black hole cases. (a): the observed intensities originated from the first (black), second (gold) and third (red) transfer function for CSH black hole case. (b): the total observed intensities $I_{obs}/I_0$ of the CSH (the red line) and RN (the blue line) black hole cases. (c): the two-dimensional disk of observed intensities for CSH and RN black hole cases with $Q=0.99M$.}\label{figprofilea}
\end{center}
\end{figure}

Figure \ref{figprofilea} shows the observed intensities and black hole images for the first model with the emission function of the accretion disk, represented by Eq. \eqref{diskprofile1}. In Fig. \ref{three ointensities}, it is evident that, for this model, there are three distinct peaks corresponding to the direct (the black line), lensed (the orange line), and photon ring (the red line) intensities, and these peaks are independent of each other. In other words, the three intensities do not exhibit any superposition effects on the overall observed intensities, a fact that can be seen from in Fig. \ref{total ointensities} with the red line.  For the charge parameter $Q=0.99M$ case, the direct intensity has a behavior with a sharp change occurring at approximately $b\sim5.397M$ due to the gravitational lensing effect. The lensed ring intensity is confined to a narrow range, with $b\in(4.499M,5.223M)$. In contrast, the photon ring intensity manifests as extremely narrow spikes at approximately $b\sim4.153M$, contributing negligibly to the total observed intensity. From Fig. \ref{total ointensities}, we can see that with a large charge parameter $Q=0.99M$ case, both the difference in the distribution of direct intensity and the brightness between the CSH and RN black hole cases become evident. Figure \ref{disksh1} shows the two-dimensional disk of observed intensities for CSH and RN black hole with $Q=0.99M$. 

\begin{figure}[htbp!]
\begin{center}
\subfigure[]{\label{maxb casea}
\includegraphics[width=0.23\textwidth]{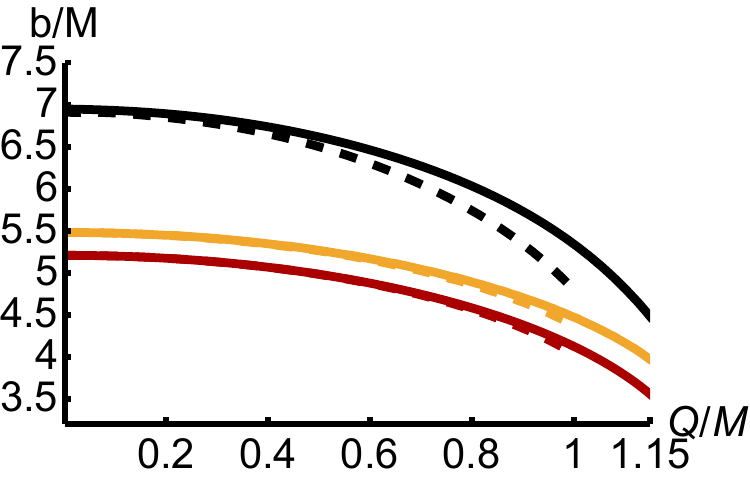}}
\subfigure[]{\label{maxI casea}
\includegraphics[width=0.23\textwidth]{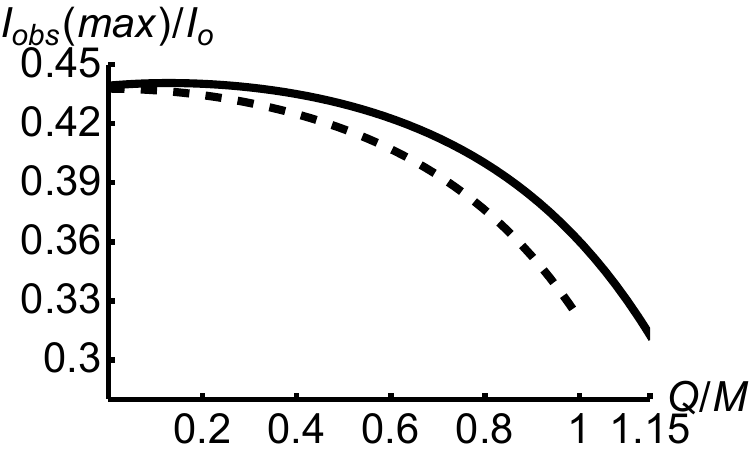}}
\caption{The impact parameter $b$ corresponding to intensity peaks and the max peak brightness of CSH (the solid lines) and RN (the dashed lines) black hole cases for \textbf{Model a}.}\label{max casea}
\end{center}
\end{figure}

Furthermore, Fig. \ref{maxb casea} illustrates the impact parameter $b$ associated with the three intensity peaks. The results indicate that discerning between the CSH and RN black hole cases based on the peaks of lensed (the orange lines) and photon ring (the red lines) intensities is challenging, whereas the peaks of direct intensity (the black lines) provides more robust difference. Besides, Figs. \ref{three ointensities} and \ref{total ointensities} also show that all three of the intensity peaks exhibit nearly identical brightness. So, the impact of the charge parameter $Q$ on the three brightness should be the same, which are shown in Fig. \ref{maxI casea}. The findings reveal that the charge parameter $Q$ has a suppressive effect on brightness, with different degrees of suppression observed in the CSH and RN black holes. Based on these distinctions, it is possible to distinguish the CSH black hole and the RN black hole by examining the distribution and peak brightness of direct intensity.

\begin{figure}[htbp!]
\begin{center}
\subfigure[]{\label{three ointensitiesb}
\includegraphics[width=0.23\textwidth]{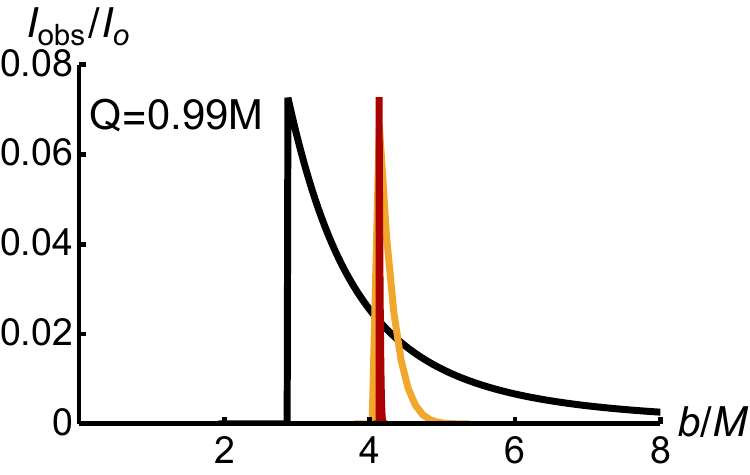}}
\subfigure[]{\label{total ointensitiesb}
\includegraphics[width=0.23\textwidth]{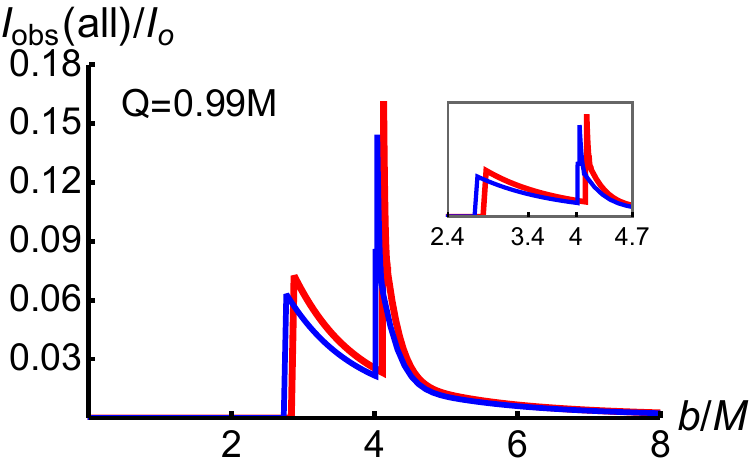}}
\subfigure[]{\label{diskshb}
\includegraphics[width=0.18\textwidth]{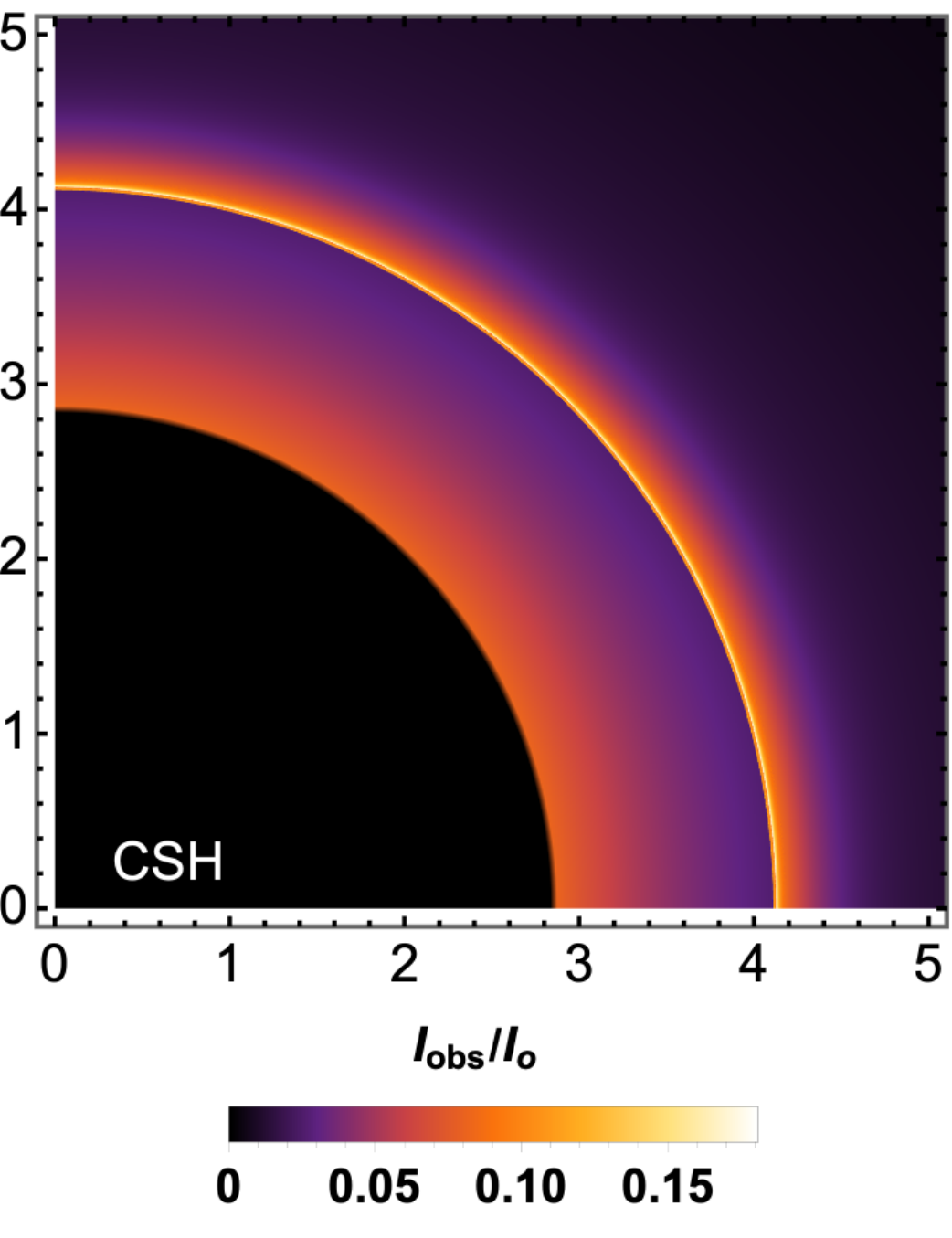}\hspace{1cm}
\includegraphics[width=0.18\textwidth]{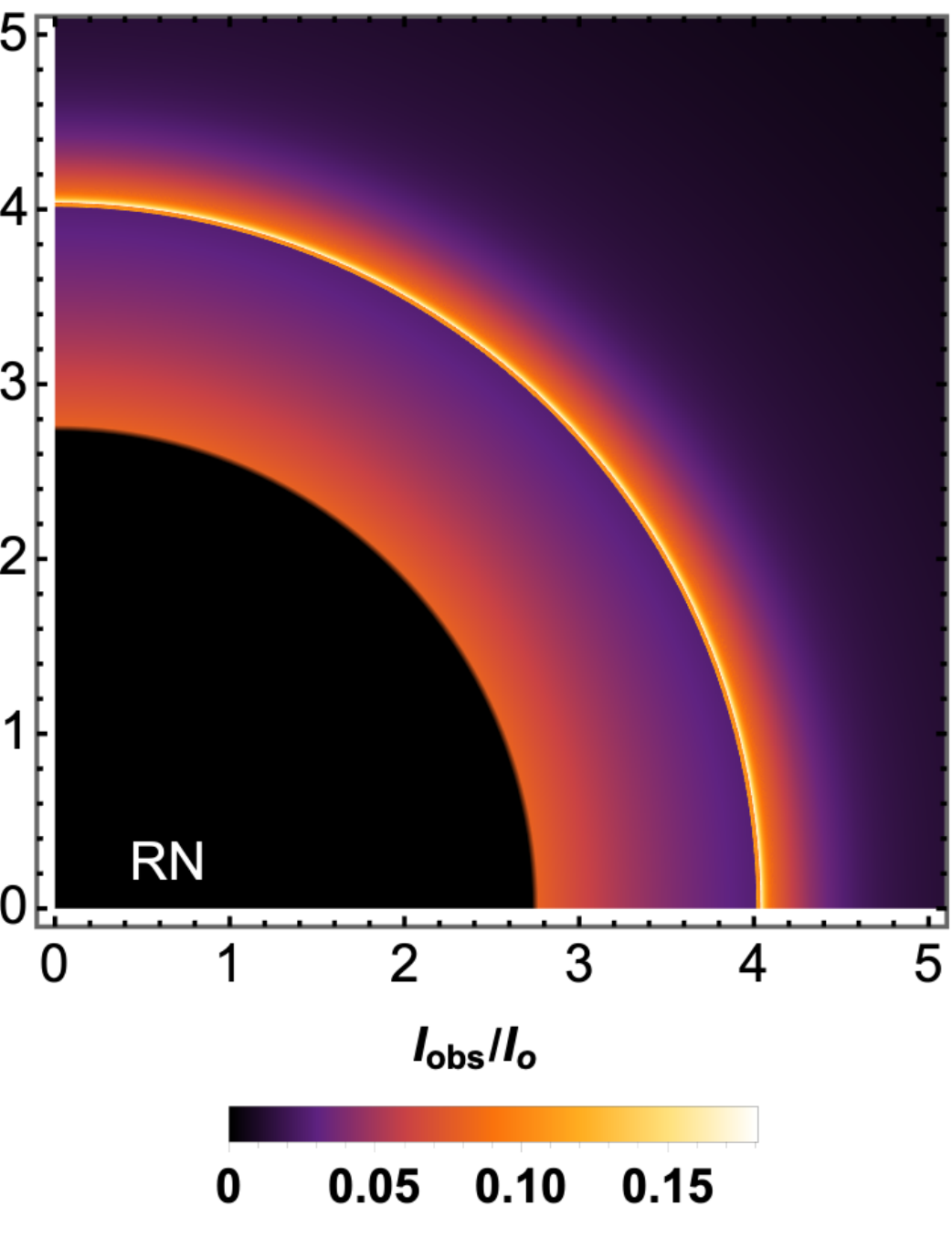}}
\caption{Observational appearances of the thin disk with \textbf{Model b} for the CSH and RN black hole cases. (a): the observed intensities originated from the first (black), second (gold) and third (red) transfer function for CSH black hole case. (b): the total observed intensities $I_{obs}/I_0$ of the CSH (the red line) and RN (the blue line) black hole cases. (c): the two-dimensional disk of observed intensities for CSH and RN black hole cases with $Q=0.99M$.}\label{figprofileb}
\end{center}
\end{figure}

\begin{figure}[htbp!]
\begin{center}
\subfigure[]{\label{maxb caseb}
\includegraphics[width=0.23\textwidth]{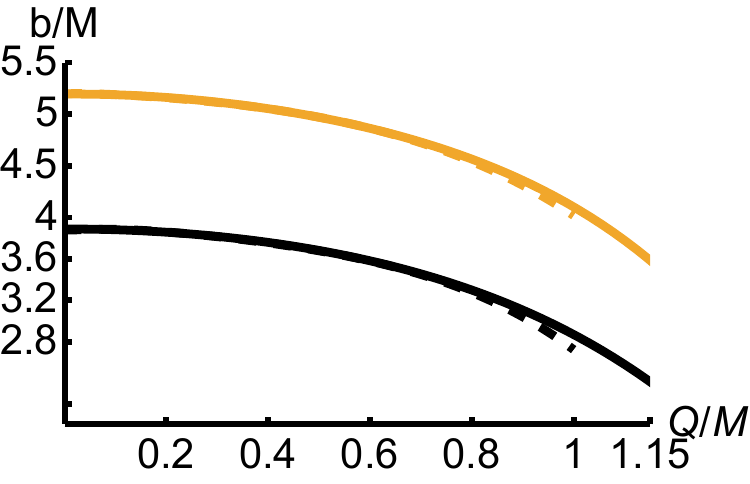}}
\subfigure[]{\label{maxI caseb}
\includegraphics[width=0.23\textwidth]{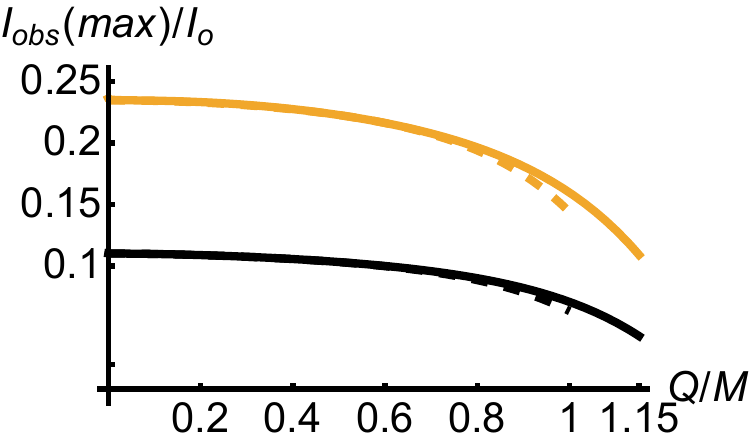}}
\caption{The impact parameter $b$ corresponding to the intensity peaks and the max peak brightness of CSH (the solid line) and RN (the dashed line) black hole cases for \textbf{Model b}.}\label{max caseb}
\end{center}
\end{figure}

{Figure \ref{figprofileb} displays the observed intensities and black hole images for \textbf{Model b}.} From Fig. \ref{three ointensitiesb}, it's apparent that in this model, the three intensities are not independent of each other. The direct intensity is superimposed by the lensed and photon ring intensities with $b\in(3.833M,5.323M)$, and exhibit abrupt changes at approximately $b\sim2.873M$ for $Q=0.99M$. For the total observed intensities shown in Fig. \ref{total ointensitiesb}, in the innermost part there is a peak which contributed only by the direct intensity. The combination of the three intensities creates a maximum peak. Consequently, there are two distinct peaks attributed to the direct intensity alone and a superimposed peak resulting from all three intensities together. The two-dimensional disks of the total observed intensities for the two black hole cases are shown in Fig. \ref{diskshb}.  Figure \ref{max caseb} also illustrates the positions of two distinct peaks with the impact parameter $b$ and the corresponding peak brightness, which are the direct intensity peaks (the black lines) and the superimposed peaks (the orange lines). The results indicate that the difference between the two peak positions corresponding to the CSH and RN black holes is not sufficiently clear.  Compared to the brightness contributed solely by the direct intensity, the superimposed brightness facilitates a more straightforward difference between CSH and RN black holes.

\begin{figure}[htbp!]
\begin{center}
\subfigure[]{\label{three ointensitiesc}
\includegraphics[width=0.23\textwidth]{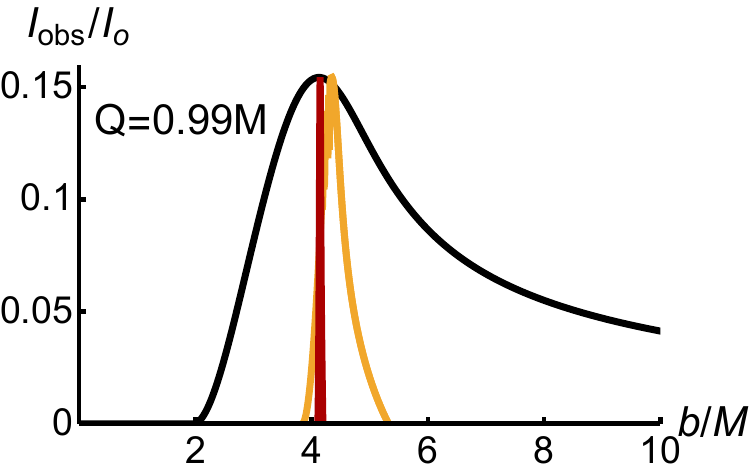}}
\subfigure[]{\label{total ointensitiesc}
\includegraphics[width=0.23\textwidth]{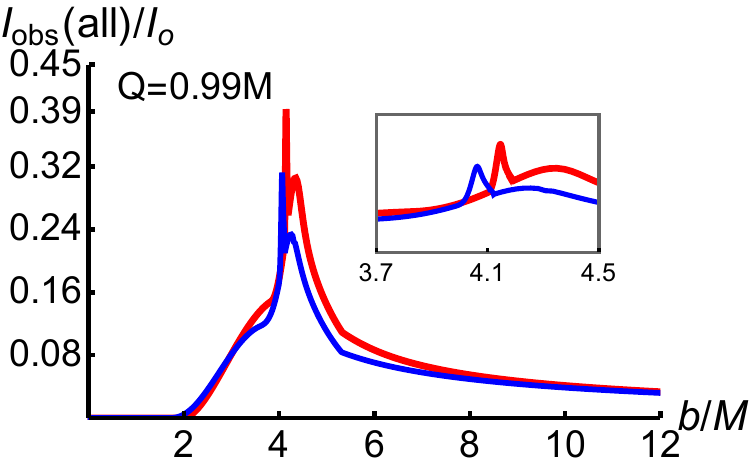}}
\subfigure[]{\label{diskshc}
\includegraphics[width=0.2\textwidth]{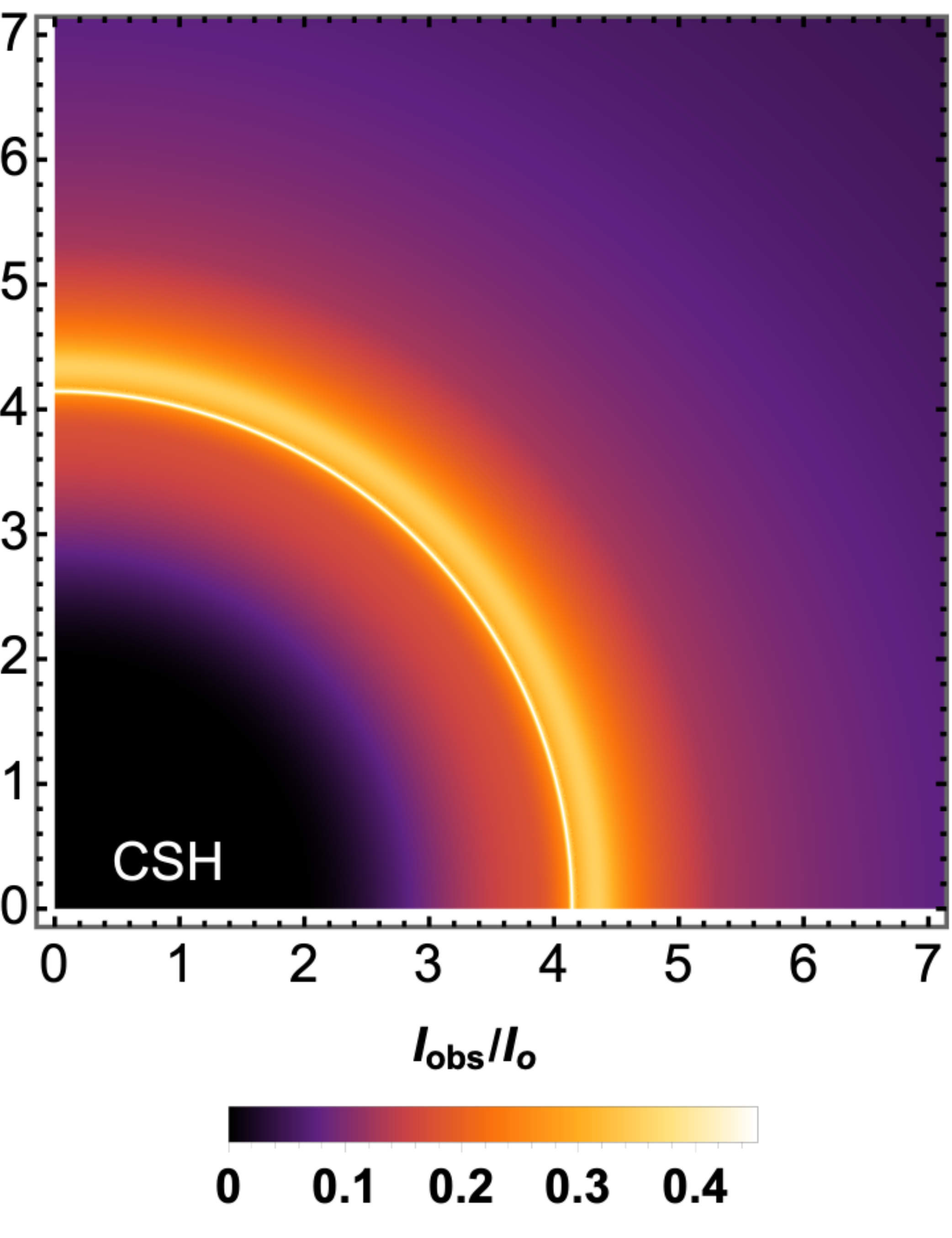}\hspace{0.8cm}
\includegraphics[width=0.2\textwidth]{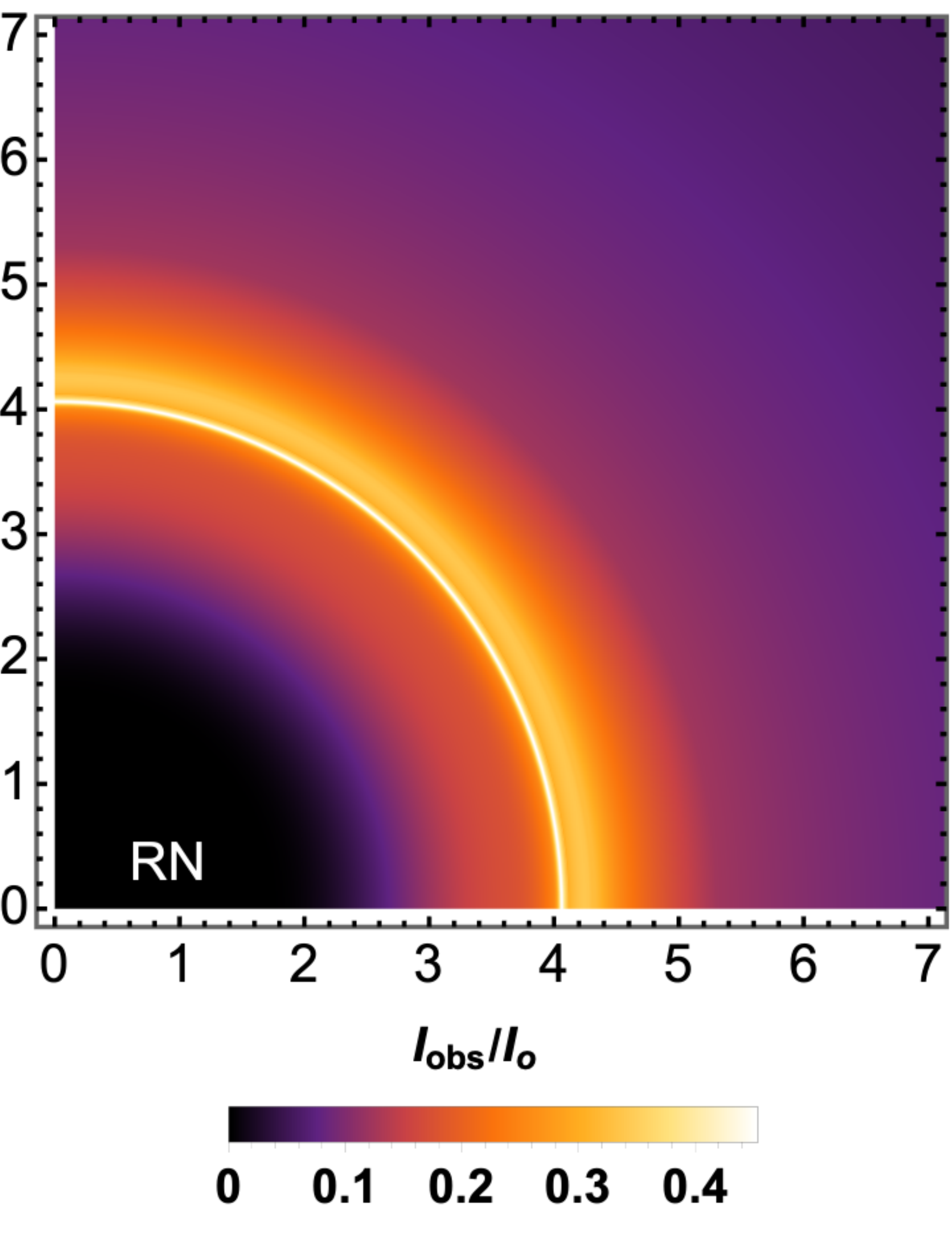}}
\caption{Observational appearances of the thin disk with \textbf{Model c} for the CSH and RN black hole cases. (a): the observed intensities originated from the first (black), second (gold) and third (red) transfer function for CSH black hole case. (b): the total observed intensities $I_{obs}/I_0$ of the CSH (the red line) and RN (the blue line) black hole cases. (c): the two-dimensional disk of observed intensities for CSH and RN black hole cases with $Q=0.99M$.}\label{figprofilec}
\end{center}
\end{figure}

\begin{figure}[htp!]
\begin{center}
\subfigure[]{\label{maxb casec}
\includegraphics[width=0.23\textwidth]{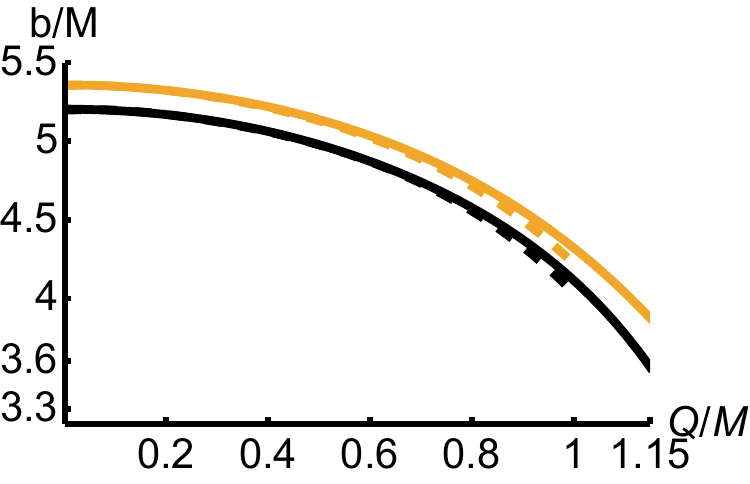}}
\subfigure[]{\label{maxI casec}
\includegraphics[width=0.23\textwidth]{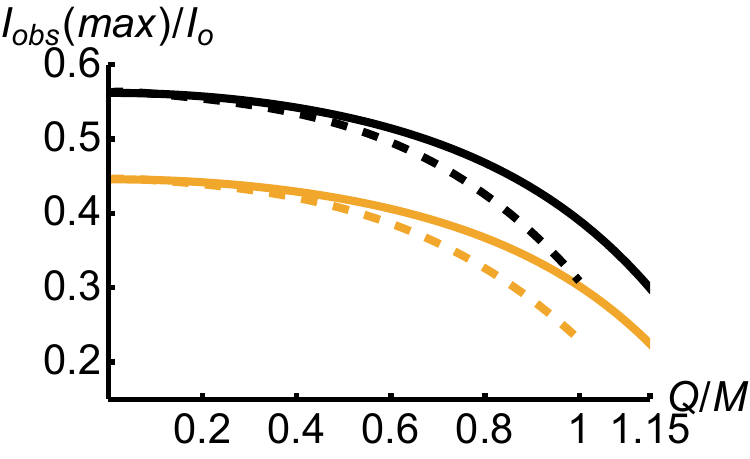}}
\caption{ The impact parameter $b$ corresponding to the intensity peaks and the max peak brightness of CSH (the solid line) and RN (the dashed line) black hole cases for \textbf{Model c} case. The colors black and orange correspond to the two peaks, arranged from the innermost to the outermost.}\label{max casec}
\end{center}
\end{figure}

{Figure \ref{figprofilec} presents the observed intensities and black hole images for the third emission function case with $Q=0.99M$.} In Fig. \ref{three ointensitiesc}, it is noticeable that compared to the previous two emission function cases, the direct intensity do not exhibit abrupt terminations at any impact parameter $b$, and begins from $b\sim2M$. The photon ring intensity remains in the middle with $b\sim4.145M$, while the lensed intensity wraps around $b\in(3.812M,5.312M)$. This implies that all the three intensities have a cumulative effect on the total observed intensities, as shown in Fig. \ref{total ointensitiesc}. The results highlight that both the photon ring and lensed ring intensities make significant contribution to the peak brightness, which means the combined effects result two peaks in the total observed intensities. The two-dimensional representations of the total observed intensities for both black holes are presented in Figs. \ref{diskshc}. Additionally, Fig. \ref{max casec} shows the positions of the two peaks along with the impact parameter $b$ and peak brightness, which are contributed by the photon intensity (the black lines) and the lensed intensity (the orange lines). The result shows that, comparing the influence of the charge parameter $Q$ on the position of peaks, the effect of $Q$ on brightness is more effective for distinguishing between CSH and RN black holes.

\section{Conclusion and discussion}\label{conclusion}

In this paper, we have undertaken an exploration of the optical characteristics of CSH black hole within the context of a four-dimensional Einstein-Maxwell-Dilaton theory, which can be derived from a five-dimensional Einstein-Maxwell theory. In this scenario, the associated electric charge range for CSH black hole is $Q\in[0,2/\sqrt{3}M]$. Our analysis primarily focused on examining the impact of the charge parameter $Q$ on photon trajectories and the black hole images when illuminated by different optical and geometrically thin accretion disks.

Initially, we delved into the trajectories of photons around the CSH black hole, particularly in the equatorial plane. We obtained the time, azimuthal, and radial components of the four-velocity. Notably, as the charge parameter $Q$ increased, the discrepancy between the effective potentials $V_{eff}$, radius of the photon sphere $r_{ph}$ and the innermost stable circular orbit $r_{isco}$ for CSH and RN black hole cases became increasingly pronounced. While, in the range of $Q\in[0,M]$, it is not possible to discern any difference between the CSH and RN black hole cases in terms of the critical impact parameter $b_{ph}$.

We also analyzed the effect of the charge parameter $Q$ on the orbit number $n$. Our findings reveal that as the charge parameter $Q$ approaches to $M$, the difference between the two black hole cases becomes more pronounced based on the orbit number $n$. To describe the relationship between the photon ring and the lensed ring bands, we introduced a ratio, $\Gamma$. It is evident that the increment of $\Gamma$ with Q is less significant for the CSH black hole than that of the  RN black hole. However, if $\Gamma$ exceeds $0.1$, it may suggest the presence of a charged black hole with scalar hair.

Then, we investigated the images of CSH and RN black hole cases, illuminated by three accretion disk toy models which are optical and geometrically thin. Our detailed analysis was focused on assessing the impact of the charge parameter $Q$ on the brightness of direct, lensed ring, and photon ring intensities, as well as their contributions to the total observed intensities.  {The findings indicate that, for \textbf{Model a}, the difference between CSH and RN black holes becomes feasible through an analysis of the distribution and the peak brightness of direct intensity. Conversely, in the context of \textbf{Model b}, the increased brightness resulting from variations in the parameter $Q$ significantly enhances the difference between CSH and RN black holes. Within the framework of \textbf{Model c}, it is demonstrated that the influence of $Q$ on brightness offers a more effective method for distinguishing between CSH and RN black holes.}

\section*{acknowledgments}
This work is supported in part by the National Natural Science Foundation of China (Grants No. 12147175, No. 12205129, No. 12147166, No. 12175105, No. 11575083, No. 11565017), the China Postdoctoral Science Foundation (Grant No. 2021M701529), and the Natural Science Foundation of Jiangsu Province (Grant No. \\BK20220642).


\begin{thebibliography}{49}%

\bibitem{SupernovaSearchTeam:1998fmf}
{\scshape Supernova Search Team} collaboration, A.~G. Riess et~al.,
  \emph{{Observational evidence from supernovae for an accelerating universe
  and a cosmological constant}},
  \href{http://dx.doi.org/10.1086/300499}{\emph{Astron. J.} {\bf 116}, (1998)
  1009--1038}, \href{http://arxiv.org/abs/astro-ph/9805201}{{\tt
  astro-ph/9805201}}.

\bibitem{SupernovaCosmologyProject:1998vns}
{\scshape Supernova Cosmology Project} collaboration, S.~Perlmutter et~al.,
  \emph{{Measurements of $\Omega$ and $\Lambda$ from 42 high redshift
  supernovae}}, \href{http://dx.doi.org/10.1086/307221}{\emph{Astrophys. J.}
  {\bf 517}, (1999) 565--586},
  \href{http://arxiv.org/abs/astro-ph/9812133}{{\tt astro-ph/9812133}}.

\bibitem{Clifton:2011jh}
T.~Clifton, P.~G. Ferreira, A.~Padilla and C.~Skordis, \emph{{Modified Gravity
  and Cosmology}},
  \href{http://dx.doi.org/10.1016/j.physrep.2012.01.001}{\emph{Phys. Rept.}
  {\bf 513}, (2012) 1--189}, \href{http://arxiv.org/abs/1106.2476}{{\tt
  arXiv:1106.2476}}.

\bibitem{Horndeski:1974wa}
G.~W. Horndeski, \emph{{Second-order scalar-tensor field equations in a
  four-dimensional space}},
  \href{http://dx.doi.org/10.1007/BF01807638}{\emph{Int. J. Theor. Phys.} {\bf
  10}, (1974) 363--384}.

\bibitem{Rinaldi:2012vy}
M.~Rinaldi, \emph{{Black holes with non-minimal derivative coupling}},
  \href{http://dx.doi.org/10.1103/PhysRevD.86.084048}{\emph{Phys. Rev. D} {\bf
  86}, (2012) 084048}, \href{http://arxiv.org/abs/1208.0103}{{\tt
  arXiv:1208.0103}}.

\bibitem{Cisterna:2014nua}
A.~Cisterna and C.~Erices, \emph{{Asymptotically locally AdS and flat black
  holes in the presence of an electric field in the Horndeski scenario}},
  \href{http://dx.doi.org/10.1103/PhysRevD.89.084038}{\emph{Phys. Rev. D} {\bf
  89}, (2014) 084038}, \href{http://arxiv.org/abs/1401.4479}{{\tt
  arXiv:1401.4479}}.

\bibitem{Feng:2015oea}
X.-H. Feng, H.-S. Liu, H.~L\"u and C.~N. Pope, \emph{{Black Hole Entropy and
  Viscosity Bound in Horndeski Gravity}},
  \href{http://dx.doi.org/10.1007/JHEP11(2015)176}{\emph{JHEP} {\bf 11}, (2015)
  176}, \href{http://arxiv.org/abs/1509.07142}{{\tt arXiv:1509.07142}}.

\bibitem{Sotiriou:2013qea}
T.~P. Sotiriou and S.-Y. Zhou, \emph{{Black hole hair in generalized
  scalar-tensor gravity}},
  \href{http://dx.doi.org/10.1103/PhysRevLett.112.251102}{\emph{Phys. Rev.
  Lett.} {\bf 112}, (2014) 251102}, \href{http://arxiv.org/abs/1312.3622}{{\tt
  arXiv:1312.3622}}.

\bibitem{Miao:2016aol}
Y.-G. Miao and Z.-M. Xu, \emph{{Thermodynamics of Horndeski black holes with
  non-minimal derivative coupling}},
  \href{http://dx.doi.org/10.1140/epjc/s10052-016-4482-1}{\emph{Eur. Phys. J.
  C} {\bf 76}, (2016) 638}, \href{http://arxiv.org/abs/1607.06629}{{\tt
  arXiv:1607.06629}}.

\bibitem{Kuang:2016edj}
X.-M. Kuang and E.~Papantonopoulos, \emph{{Building a Holographic
  Superconductor with a Scalar Field Coupled Kinematically to Einstein
  Tensor}}, \href{http://dx.doi.org/10.1007/JHEP08(2016)161}{\emph{JHEP} {\bf
  08}, (2016) 161}, \href{http://arxiv.org/abs/1607.04928}{{\tt
  arXiv:1607.04928}}.

\bibitem{Babichev:2016rlq}
E.~Babichev, C.~Charmousis and A.~Leh\'ebel, \emph{{Black holes and stars in
  Horndeski theory}},
  \href{http://dx.doi.org/10.1088/0264-9381/33/15/154002}{\emph{Class. Quant.
  Grav.} {\bf 33}, (2016) 154002}, \href{http://arxiv.org/abs/1604.06402}{{\tt
  arXiv:1604.06402}}.

\bibitem{Benkel:2016rlz}
R.~Benkel, T.~P. Sotiriou and H.~Witek, \emph{{Black hole hair formation in
  shift-symmetric generalised scalar-tensor gravity}},
  \href{http://dx.doi.org/10.1088/1361-6382/aa5ce7}{\emph{Class. Quant. Grav.}
  {\bf 34}, (2017) 064001}, \href{http://arxiv.org/abs/1610.09168}{{\tt
  arXiv:1610.09168}}.

\bibitem{Filios:2018xvy}
G.~Filios, P.~A. Gonz\'alez, X.-M. Kuang, E.~Papantonopoulos and Y.~V\'asquez,
  \emph{{Spontaneous Momentum Dissipation and Coexistence of Phases in
  Holographic Horndeski Theory}},
  \href{http://dx.doi.org/10.1103/PhysRevD.99.046017}{\emph{Phys. Rev. D} {\bf
  99}, (2019) 046017}, \href{http://arxiv.org/abs/1808.07766}{{\tt
  arXiv:1808.07766}}.

\bibitem{Cisterna:2018hzf}
A.~Cisterna, C.~Erices, X.-M. Kuang and M.~Rinaldi, \emph{{Axionic black branes
  with conformal coupling}},
  \href{http://dx.doi.org/10.1103/PhysRevD.97.124052}{\emph{Phys. Rev. D} {\bf
  97}, (2018) 124052}, \href{http://arxiv.org/abs/1803.07600}{{\tt
  arXiv:1803.07600}}.

\bibitem{Giusti:2021sku}
A.~Giusti, S.~Zentarra, L.~Heisenberg and V.~Faraoni, \emph{{First-order
  thermodynamics of Horndeski gravity}},
  \href{http://dx.doi.org/10.1103/PhysRevD.105.124011}{\emph{Phys. Rev. D} {\bf
  105}, (2022) 124011}, \href{http://arxiv.org/abs/2108.10706}{{\tt
  arXiv:2108.10706}}.

\bibitem{Babichev:2013cya}
E.~Babichev and C.~Charmousis, \emph{{Dressing a black hole with a
  time-dependent Galileon}},
  \href{http://dx.doi.org/10.1007/JHEP08(2014)106}{\emph{JHEP} {\bf 08}, (2014)
  106}, \href{http://arxiv.org/abs/1312.3204}{{\tt arXiv:1312.3204}}.

\bibitem{Babichev:2017lmw}
E.~Babichev, C.~Charmousis, G.~Esposito-Far\`ese and A.~Leh\'ebel,
  \emph{{Stability of Black Holes and the Speed of Gravitational Waves within
  Self-Tuning Cosmological Models}},
  \href{http://dx.doi.org/10.1103/PhysRevLett.120.241101}{\emph{Phys. Rev.
  Lett.} {\bf 120}, (2018) 241101}, \href{http://arxiv.org/abs/1712.04398}{{\tt
  arXiv:1712.04398}}.

\bibitem{BenAchour:2018dap}
J.~Ben~Achour and H.~Liu, \emph{{Hairy Schwarzschild-(A)dS black hole solutions
  in degenerate higher order scalar-tensor theories beyond shift symmetry}},
  \href{http://dx.doi.org/10.1103/PhysRevD.99.064042}{\emph{Phys. Rev. D} {\bf
  99}, (2019) 064042}, \href{http://arxiv.org/abs/1811.05369}{{\tt
  arXiv:1811.05369}}.

\bibitem{Takahashi:2019oxz}
K.~Takahashi, H.~Motohashi and M.~Minamitsuji, \emph{{Linear stability analysis
  of hairy black holes in quadratic degenerate higher-order scalar-tensor
  theories: Odd-parity perturbations}},
  \href{http://dx.doi.org/10.1103/PhysRevD.100.024041}{\emph{Phys. Rev. D} {\bf
  100}, (2019) 024041}, \href{http://arxiv.org/abs/1904.03554}{{\tt
  arXiv:1904.03554}}.

\bibitem{Minamitsuji:2019shy}
M.~Minamitsuji and J.~Edholm, \emph{{Black hole solutions in shift-symmetric
  degenerate higher-order scalar-tensor theories}},
  \href{http://dx.doi.org/10.1103/PhysRevD.100.044053}{\emph{Phys. Rev. D} {\bf
  100}, (2019) 044053}, \href{http://arxiv.org/abs/1907.02072}{{\tt
  arXiv:1907.02072}}.

\bibitem{Arkani-Hamed:2003juy}
N.~Arkani-Hamed, P.~Creminelli, S.~Mukohyama and M.~Zaldarriaga, \emph{{Ghost
  inflation}},
  \href{http://dx.doi.org/10.1088/1475-7516/2004/04/001}{\emph{JCAP} {\bf 04},
  (2004) 001}, \href{http://arxiv.org/abs/hep-th/0312100}{{\tt
  hep-th/0312100}}.

\bibitem{Bah:2020pdz}
I.~Bah and P.~Heidmann, \emph{{Topological stars, black holes and generalized
  charged Weyl solutions}},
  \href{http://dx.doi.org/10.1007/JHEP09(2021)147}{\emph{JHEP} {\bf 09}, (2021)
  147}, \href{http://arxiv.org/abs/2012.13407}{{\tt arXiv:2012.13407}}.

\bibitem{Bah:2020ogh}
I.~Bah and P.~Heidmann, \emph{{Topological Stars and Black Holes}},
  \href{http://dx.doi.org/10.1103/PhysRevLett.126.151101}{\emph{Phys. Rev.
  Lett.} {\bf 126}, (2021) 151101}, \href{http://arxiv.org/abs/2011.08851}{{\tt
  arXiv:2011.08851}}.

\bibitem{Bah:2022yji}
I.~Bah, P.~Heidmann and P.~Weck, \emph{{Schwarzschild-like topological
  solitons}}, \href{http://dx.doi.org/10.1007/JHEP08(2022)269}{\emph{JHEP} {\bf
  08}, (2022) 269}, \href{http://arxiv.org/abs/2203.12625}{{\tt
  arXiv:2203.12625}}.

\bibitem{Stotyn:2011tv}
S.~Stotyn and R.~B. Mann, \emph{{Magnetic charge can locally stabilize
  Kaluza\textendash{}Klein bubbles}},
  \href{http://dx.doi.org/10.1016/j.physletb.2011.10.015}{\emph{Phys. Lett. B}
  {\bf 705}, (2011) 269--272}, \href{http://arxiv.org/abs/1105.1854}{{\tt
  arXiv:1105.1854}}.

\bibitem{Heidmann:2022ehn}
P.~Heidmann, I.~Bah and E.~Berti, \emph{{Imaging topological solitons: The
  microstructure behind the shadow}},
  \href{http://dx.doi.org/10.1103/PhysRevD.107.084042}{\emph{Phys. Rev. D} {\bf
  107}, (2023) 084042}, \href{http://arxiv.org/abs/2212.06837}{{\tt
  arXiv:2212.06837}}.

\bibitem{Guo:2022nto}
W.-D. Guo, S.-W. Wei and Y.-X. Liu, \emph{{Shadow of a charged black hole with
  scalar hair}},
  \href{http://dx.doi.org/10.1140/epjc/s10052-023-11316-1}{\emph{Eur. Phys. J.
  C} {\bf 83}, (2023) 197}, \href{http://arxiv.org/abs/2203.13477}{{\tt
  arXiv:2203.13477}}.

\bibitem{Guo:2022rms}
W.-D. Guo, Q.~Tan and Y.-X. Liu, \emph{{Gravitoelectromagnetic coupled
  perturbations and quasinormal modes of a charged black hole with scalar
  hair}}, \href{http://dx.doi.org/10.1103/PhysRevD.107.124046}{\emph{Phys. Rev.
  D} {\bf 107}, (2023) 124046}, \href{http://arxiv.org/abs/2212.08784}{{\tt
  arXiv:2212.08784}}.

\bibitem{EventHorizonTelescope:2019dse}
{\scshape Event Horizon Telescope} collaboration, K.~Akiyama et~al.,
  \emph{{First M87 Event Horizon Telescope Results. I. The Shadow of the
  Supermassive Black Hole}},
  \href{http://dx.doi.org/10.3847/2041-8213/ab0ec7}{\emph{Astrophys. J. Lett.}
  {\bf 875}, (2019) L1}, \href{http://arxiv.org/abs/1906.11238}{{\tt
  arXiv:1906.11238}}.

\bibitem{EventHorizonTelescope:2019uob}
{\scshape Event Horizon Telescope} collaboration, K.~Akiyama et~al.,
  \emph{{First M87 Event Horizon Telescope Results. II. Array and
  Instrumentation}},
  \href{http://dx.doi.org/10.3847/2041-8213/ab0c96}{\emph{Astrophys. J. Lett.}
  {\bf 875}, (2019) L2}, \href{http://arxiv.org/abs/1906.11239}{{\tt
  arXiv:1906.11239}}.

\bibitem{EventHorizonTelescope:2019jan}
{\scshape Event Horizon Telescope} collaboration, K.~Akiyama et~al.,
  \emph{{First M87 Event Horizon Telescope Results. III. Data Processing and
  Calibration}},
  \href{http://dx.doi.org/10.3847/2041-8213/ab0c57}{\emph{Astrophys. J. Lett.}
  {\bf 875}, (2019) L3}, \href{http://arxiv.org/abs/1906.11240}{{\tt
  arXiv:1906.11240}}.

\bibitem{EventHorizonTelescope:2019ths}
{\scshape Event Horizon Telescope} collaboration, K.~Akiyama et~al.,
  \emph{{First M87 Event Horizon Telescope Results. IV. Imaging the Central
  Supermassive Black Hole}},
  \href{http://dx.doi.org/10.3847/2041-8213/ab0e85}{\emph{Astrophys. J. Lett.}
  {\bf 875}, (2019) L4}, \href{http://arxiv.org/abs/1906.11241}{{\tt
  arXiv:1906.11241}}.

\bibitem{EventHorizonTelescope:2019pgp}
{\scshape Event Horizon Telescope} collaboration, K.~Akiyama et~al.,
  \emph{{First M87 Event Horizon Telescope Results. V. Physical Origin of the
  Asymmetric Ring}},
  \href{http://dx.doi.org/10.3847/2041-8213/ab0f43}{\emph{Astrophys. J. Lett.}
  {\bf 875}, (2019) L5}, \href{http://arxiv.org/abs/1906.11242}{{\tt
  arXiv:1906.11242}}.

\bibitem{EventHorizonTelescope:2019ggy}
{\scshape Event Horizon Telescope} collaboration, K.~Akiyama et~al.,
  \emph{{First M87 Event Horizon Telescope Results. VI. The Shadow and Mass of
  the Central Black Hole}},
  \href{http://dx.doi.org/10.3847/2041-8213/ab1141}{\emph{Astrophys. J. Lett.}
  {\bf 875}, (2019) L6}, \href{http://arxiv.org/abs/1906.11243}{{\tt
  arXiv:1906.11243}}.

\bibitem{EventHorizonTelescope:2022wkp}
{\scshape Event Horizon Telescope} collaboration, K.~Akiyama et~al.,
  \emph{{First Sagittarius A* Event Horizon Telescope Results. I. The Shadow of
  the Supermassive Black Hole in the Center of the Milky Way}},
  \href{http://dx.doi.org/10.3847/2041-8213/ac6674}{\emph{Astrophys. J. Lett.}
  {\bf 930}, (2022) L12}.

\bibitem{EventHorizonTelescope:2022apq}
{\scshape Event Horizon Telescope} collaboration, K.~Akiyama et~al.,
  \emph{{First Sagittarius A* Event Horizon Telescope Results. II. EHT and
  Multiwavelength Observations, Data Processing, and Calibration}},
  \href{http://dx.doi.org/10.3847/2041-8213/ac6675}{\emph{Astrophys. J. Lett.}
  {\bf 930}, (2022) L13}.

\bibitem{EventHorizonTelescope:2022wok}
{\scshape Event Horizon Telescope} collaboration, K.~Akiyama et~al.,
  \emph{{First Sagittarius A* Event Horizon Telescope Results. III. Imaging of
  the Galactic Center Supermassive Black Hole}},
  \href{http://dx.doi.org/10.3847/2041-8213/ac6429}{\emph{Astrophys. J. Lett.}
  {\bf 930}, (2022) L14}.

\bibitem{EventHorizonTelescope:2022exc}
{\scshape Event Horizon Telescope} collaboration, K.~Akiyama et~al.,
  \emph{{First Sagittarius A* Event Horizon Telescope Results. IV. Variability,
  Morphology, and Black Hole Mass}},
  \href{http://dx.doi.org/10.3847/2041-8213/ac6736}{\emph{Astrophys. J. Lett.}
  {\bf 930}, (2022) L15}.

\bibitem{EventHorizonTelescope:2022urf}
{\scshape Event Horizon Telescope} collaboration, K.~Akiyama et~al.,
  \emph{{First Sagittarius A* Event Horizon Telescope Results. V. Testing
  Astrophysical Models of the Galactic Center Black Hole}},
  \href{http://dx.doi.org/10.3847/2041-8213/ac6672}{\emph{Astrophys. J. Lett.}
  {\bf 930}, (2022) L16}.

\bibitem{EventHorizonTelescope:2022xqj}
{\scshape Event Horizon Telescope} collaboration, K.~Akiyama et~al.,
  \emph{{First Sagittarius A* Event Horizon Telescope Results. VI. Testing the
  Black Hole Metric}},
  \href{http://dx.doi.org/10.3847/2041-8213/ac6756}{\emph{Astrophys. J. Lett.}
  {\bf 930}, (2022) L17}.

\bibitem{Kumar:2018ple}
R.~Kumar and S.~G. Ghosh, \emph{{Black Hole Parameter Estimation from Its
  Shadow}}, \href{http://dx.doi.org/10.3847/1538-4357/ab77b0}{\emph{Astrophys.
  J.} {\bf 892}, (2020) 78}, \href{http://arxiv.org/abs/1811.01260}{{\tt
  arXiv:1811.01260}}.

\bibitem{Ghosh:2020spb}
S.~G. Ghosh, R.~Kumar and S.~U. Islam, \emph{{Parameters estimation and strong
  gravitational lensing of nonsingular Kerr-Sen black holes}},
  \href{http://dx.doi.org/10.1088/1475-7516/2021/03/056}{\emph{JCAP} {\bf 03},
  (2021) 056}, \href{http://arxiv.org/abs/2011.08023}{{\tt arXiv:2011.08023}}.

\bibitem{Afrin:2021imp}
M.~Afrin, R.~Kumar and S.~G. Ghosh, \emph{{Parameter estimation of hairy Kerr
  black holes from its shadow and constraints from M87*}},
  \href{http://dx.doi.org/10.1093/mnras/stab1260}{\emph{Mon. Not. Roy. Astron.
  Soc.} {\bf 504}, (2021) 5927--5940},
  \href{http://arxiv.org/abs/2103.11417}{{\tt arXiv:2103.11417}}.

\bibitem{Ghosh:2022kit}
S.~G. Ghosh and M.~Afrin, \emph{{An Upper Limit on the Charge of the Black Hole
  Sgr A* from EHT Observations}},
  \href{http://dx.doi.org/10.3847/1538-4357/acb695}{\emph{Astrophys. J.} {\bf
  944}, (2023) 174}, \href{http://arxiv.org/abs/2206.02488}{{\tt
  arXiv:2206.02488}}.

\bibitem{Vagnozzi:2019apd}
S.~Vagnozzi and L.~Visinelli, \emph{{Hunting for extra dimensions in the shadow
  of M87*}}, \href{http://dx.doi.org/10.1103/PhysRevD.100.024020}{\emph{Phys.
  Rev. D} {\bf 100}, (2019) 024020},
  \href{http://arxiv.org/abs/1905.12421}{{\tt arXiv:1905.12421}}.

\bibitem{Banerjee:2019nnj}
I.~Banerjee, S.~Chakraborty and S.~SenGupta, \emph{{Silhouette of M87*: A New
  Window to Peek into the World of Hidden Dimensions}},
  \href{http://dx.doi.org/10.1103/PhysRevD.101.041301}{\emph{Phys. Rev. D} {\bf
  101}, (2020) 041301}, \href{http://arxiv.org/abs/1909.09385}{{\tt
  arXiv:1909.09385}}.

\bibitem{Tang:2022hsu}
Z.-Y. Tang, X.-M. Kuang, B.~Wang and W.-L. Qian, \emph{{The length of a compact
  extra dimension from black hole shadow}},
  \href{http://dx.doi.org/10.1016/j.scib.2022.11.002}{\emph{Sci. Bull.} {\bf
  67}, (2022) 2272--2275}, \href{http://arxiv.org/abs/2206.08608}{{\tt
  arXiv:2206.08608}}.

\bibitem{Mizuno:2018lxz}
Y.~Mizuno, Z.~Younsi, C.~M. Fromm, O.~Porth, M.~De~Laurentis, H.~Olivares
  et~al., \emph{{The Current Ability to Test Theories of Gravity with Black
  Hole Shadows}},
  \href{http://dx.doi.org/10.1038/s41550-018-0449-5}{\emph{Nature Astron.} {\bf
  2}, (2018) 585--590}, \href{http://arxiv.org/abs/1804.05812}{{\tt
  arXiv:1804.05812}}.

\bibitem{Psaltis:2018xkc}
D.~Psaltis, \emph{{Testing General Relativity with the Event Horizon
  Telescope}}, \href{http://dx.doi.org/10.1007/s10714-019-2611-5}{\emph{Gen.
  Rel. Grav.} {\bf 51}, (2019) 137},
  \href{http://arxiv.org/abs/1806.09740}{{\tt arXiv:1806.09740}}.

\bibitem{Stepanian:2021vvk}
A.~Stepanian, S.~Khlghatyan and V.~G. Gurzadyan, \emph{{Black hole shadow to
  probe modified gravity}},
  \href{http://dx.doi.org/10.1140/epjp/s13360-021-01119-2}{\emph{Eur. Phys. J.
  Plus} {\bf 136}, (2021) 127}, \href{http://arxiv.org/abs/2101.08261}{{\tt
  arXiv:2101.08261}}.

\bibitem{Younsi:2021dxe}
Z.~Younsi, D.~Psaltis and F.~\"Ozel, \emph{{Black Hole Images as Tests of
  General Relativity: Effects of Spacetime Geometry}},
  \href{http://dx.doi.org/10.3847/1538-4357/aca58a}{\emph{Astrophys. J.} {\bf
  942}, (2023) 47}, \href{http://arxiv.org/abs/2111.01752}{{\tt
  arXiv:2111.01752}}.

\bibitem{KumarWalia:2022aop}
R.~Kumar~Walia, S.~G. Ghosh and S.~D. Maharaj, \emph{{Testing Rotating Regular
  Metrics with EHT Results of Sgr A*}},
  \href{http://dx.doi.org/10.3847/1538-4357/ac9623}{\emph{Astrophys. J.} {\bf
  939}, (2022) 77}, \href{http://arxiv.org/abs/2207.00078}{{\tt
  arXiv:2207.00078}}.

\bibitem{Vagnozzi:2022moj}
S.~Vagnozzi et~al., \emph{{Horizon-scale tests of gravity theories and
  fundamental physics from the Event Horizon Telescope image of Sagittarius
  A$^*$}},  \href{http://arxiv.org/abs/2205.07787}{{\tt arXiv:2205.07787}}.

\bibitem{Meng:2022kjs}
Y.~Meng, X.-M. Kuang and Z.-Y. Tang, \emph{{Photon regions, shadow observables,
  and constraints from M87* of a charged rotating black hole}},
  \href{http://dx.doi.org/10.1103/PhysRevD.106.064006}{\emph{Phys. Rev. D} {\bf
  106}, (2022) 064006}, \href{http://arxiv.org/abs/2204.00897}{{\tt
  arXiv:2204.00897}}.

\bibitem{Kuang:2022ojj}
X.-M. Kuang, Z.-Y. Tang, B.~Wang and A.~Wang, \emph{{Constraining a modified
  gravity theory in strong gravitational lensing and black hole shadow
  observations}},
  \href{http://dx.doi.org/10.1103/PhysRevD.106.064012}{\emph{Phys. Rev. D} {\bf
  106}, (2022) 064012}, \href{http://arxiv.org/abs/2206.05878}{{\tt
  arXiv:2206.05878}}.

\bibitem{Gussmann:2021mjj}
A.~Gu\ss{}mann, \emph{{Polarimetric signatures of the photon ring of a black
  hole that is pierced by a cosmic axion string}},
  \href{http://dx.doi.org/10.1007/JHEP08(2021)160}{\emph{JHEP} {\bf 08}, (2021)
  160}, \href{http://arxiv.org/abs/2105.06659}{{\tt arXiv:2105.06659}}.

\bibitem{Khodadi:2022pqh}
M.~Khodadi and G.~Lambiase, \emph{{Probing Lorentz symmetry violation using the
  first image of Sagittarius A*: Constraints on standard-model extension
  coefficients}},
  \href{http://dx.doi.org/10.1103/PhysRevD.106.104050}{\emph{Phys. Rev. D} {\bf
  106}, (2022) 104050}, \href{http://arxiv.org/abs/2206.08601}{{\tt
  arXiv:2206.08601}}.

\bibitem{Khodadi:2021gbc}
M.~Khodadi, G.~Lambiase and D.~F. Mota, \emph{{No-hair theorem in the wake of
  Event Horizon Telescope}},
  \href{http://dx.doi.org/10.1088/1475-7516/2021/09/028}{\emph{JCAP} {\bf 09},
  (2021) 028}, \href{http://arxiv.org/abs/2107.00834}{{\tt arXiv:2107.00834}}.

\bibitem{Schunck:2003kk}
F.~E. Schunck and E.~W. Mielke, \emph{{General relativistic boson stars}},
  \href{http://dx.doi.org/10.1088/0264-9381/20/20/201}{\emph{Class. Quant.
  Grav.} {\bf 20}, (2003) R301--R356},
  \href{http://arxiv.org/abs/0801.0307}{{\tt arXiv:0801.0307}}.

\bibitem{Hawking:2016msc}
S.~W. Hawking, M.~J. Perry and A.~Strominger, \emph{{Soft Hair on Black
  Holes}}, \href{http://dx.doi.org/10.1103/PhysRevLett.116.231301}{\emph{Phys.
  Rev. Lett.} {\bf 116}, (2016) 231301},
  \href{http://arxiv.org/abs/1601.00921}{{\tt arXiv:1601.00921}}.

\bibitem{Mazur:2004fk}
P.~O. Mazur and E.~Mottola, \emph{{Gravitational vacuum condensate stars}},
  \href{http://dx.doi.org/10.1073/pnas.0402717101}{\emph{Proc. Nat. Acad. Sci.}
  {\bf 101}, (2004) 9545--9550}, \href{http://arxiv.org/abs/gr-qc/0407075}{{\tt
  gr-qc/0407075}}.

\bibitem{Mathur:2005zp}
S.~D. Mathur, \emph{{The Fuzzball proposal for black holes: An Elementary
  review}}, \href{http://dx.doi.org/10.1002/prop.200410203}{\emph{Fortsch.
  Phys.} {\bf 53}, (2005) 793--827},
  \href{http://arxiv.org/abs/hep-th/0502050}{{\tt hep-th/0502050}}.

\bibitem{Luminet:1979nyg}
J.~P. Luminet, \emph{{Image of a spherical black hole with thin accretion
  disk}}, {\emph{Astron. Astrophys.} {\bf 75}, (1979) 228--235}.

\bibitem{Bambi:2013nla}
C.~Bambi, \emph{{Can the supermassive objects at the centers of galaxies be
  traversable wormholes? The first test of strong gravity for mm/sub-mm very
  long baseline interferometry facilities}},
  \href{http://dx.doi.org/10.1103/PhysRevD.87.107501}{\emph{Phys. Rev. D} {\bf
  87}, (2013) 107501}, \href{http://arxiv.org/abs/1304.5691}{{\tt
  arXiv:1304.5691}}.

\bibitem{Gralla:2019xty}
S.~E. Gralla, D.~E. Holz and R.~M. Wald, \emph{{Black Hole Shadows, Photon
  Rings, and Lensing Rings}},
  \href{http://dx.doi.org/10.1103/PhysRevD.100.024018}{\emph{Phys. Rev. D} {\bf
  100}, (2019) 024018}, \href{http://arxiv.org/abs/1906.00873}{{\tt
  arXiv:1906.00873}}.

\bibitem{Falcke:1999pj}
H.~Falcke, F.~Melia and E.~Agol, \emph{{Viewing the shadow of the black hole at
  the galactic center}},
  \href{http://dx.doi.org/10.1086/312423}{\emph{Astrophys. J. Lett.} {\bf 528},
  (2000) L13}, \href{http://arxiv.org/abs/astro-ph/9912263}{{\tt
  astro-ph/9912263}}.

\bibitem{Narayan:2019imo}
R.~Narayan, M.~D. Johnson and C.~F. Gammie, \emph{{The Shadow of a Spherically
  Accreting Black Hole}},
  \href{http://dx.doi.org/10.3847/2041-8213/ab518c}{\emph{Astrophys. J. Lett.}
  {\bf 885}, (2019) L33}, \href{http://arxiv.org/abs/1910.02957}{{\tt
  arXiv:1910.02957}}.

\bibitem{Zeng:2020dco}
X.-X. Zeng, H.-Q. Zhang and H.~Zhang, \emph{{Shadows and photon spheres with
  spherical accretions in the four-dimensional Gauss\textendash{}Bonnet black
  hole}}, \href{http://dx.doi.org/10.1140/epjc/s10052-020-08449-y}{\emph{Eur.
  Phys. J. C} {\bf 80}, (2020) 872},
  \href{http://arxiv.org/abs/2004.12074}{{\tt arXiv:2004.12074}}.

\bibitem{Zeng:2020vsj}
X.-X. Zeng and H.-Q. Zhang, \emph{{Influence of quintessence dark energy on the
  shadow of black hole}},
  \href{http://dx.doi.org/10.1140/epjc/s10052-020-08656-7}{\emph{Eur. Phys. J.
  C} {\bf 80}, (2020) 1058}, \href{http://arxiv.org/abs/2007.06333}{{\tt
  arXiv:2007.06333}}.

\bibitem{Peng:2020wun}
J.~Peng, M.~Guo and X.-H. Feng, \emph{{Influence of quantum correction on black
  hole shadows, photon rings, and lensing rings}},
  \href{http://dx.doi.org/10.1088/1674-1137/ac06bb}{\emph{Chin. Phys. C} {\bf
  45}, (2021) 085103}, \href{http://arxiv.org/abs/2008.00657}{{\tt
  arXiv:2008.00657}}.

\bibitem{Saurabh:2020zqg}
K.~Saurabh and K.~Jusufi, \emph{{Imprints of dark matter on black hole shadows
  using spherical accretions}},
  \href{http://dx.doi.org/10.1140/epjc/s10052-021-09280-9}{\emph{Eur. Phys. J.
  C} {\bf 81}, (2021) 490}, \href{http://arxiv.org/abs/2009.10599}{{\tt
  arXiv:2009.10599}}.

\bibitem{Qin:2020xzu}
X.~Qin, S.~Chen and J.~Jing, \emph{{Image of a regular phantom compact object
  and its luminosity under spherical accretions}},
  \href{http://dx.doi.org/10.1088/1361-6382/abf712}{\emph{Class. Quant. Grav.}
  {\bf 38}, (2021) 115008}, \href{http://arxiv.org/abs/2011.04310}{{\tt
  arXiv:2011.04310}}.

\bibitem{Gan:2021pwu}
Q.~Gan, P.~Wang, H.~Wu and H.~Yang, \emph{{Photon spheres and spherical
  accretion image of a hairy black hole}},
  \href{http://dx.doi.org/10.1103/PhysRevD.104.024003}{\emph{Phys. Rev. D} {\bf
  104}, (2021) 024003}, \href{http://arxiv.org/abs/2104.08703}{{\tt
  arXiv:2104.08703}}.

\bibitem{Okyay:2021nnh}
M.~Okyay and A.~\"Ovg\"un, \emph{{Nonlinear electrodynamics effects on the
  black hole shadow, deflection angle, quasinormal modes and greybody
  factors}}, \href{http://dx.doi.org/10.1088/1475-7516/2022/01/009}{\emph{JCAP}
  {\bf 01}, (2022) 009}, \href{http://arxiv.org/abs/2108.07766}{{\tt
  arXiv:2108.07766}}.

\bibitem{Li:2021ypw}
G.-P. Li and K.-J. He, \emph{{Observational appearances of a f(R) global
  monopole black hole illuminated by various accretions}},
  \href{http://dx.doi.org/10.1140/epjc/s10052-021-09817-y}{\emph{Eur. Phys. J.
  C} {\bf 81}, (2021) 1018}.

\bibitem{Li:2021riw}
G.-P. Li and K.-J. He, \emph{{Shadows and rings of the Kehagias-Sfetsos black
  hole surrounded by thin disk accretion}},
  \href{http://dx.doi.org/10.1088/1475-7516/2021/06/037}{\emph{JCAP} {\bf 06},
  (2021) 037}, \href{http://arxiv.org/abs/2105.08521}{{\tt arXiv:2105.08521}}.

\bibitem{Guo:2021bhr}
S.~Guo, G.-R. Li and E.-W. Liang, \emph{{Influence of accretion flow and
  magnetic charge on the observed shadows and rings of the Hayward black
  hole}}, \href{http://dx.doi.org/10.1103/PhysRevD.105.023024}{\emph{Phys. Rev.
  D} {\bf 105}, (2022) 023024}, \href{http://arxiv.org/abs/2112.11227}{{\tt
  arXiv:2112.11227}}.

\bibitem{Hu:2022lek}
S.~Hu, C.~Deng, D.~Li, X.~Wu and E.~Liang, \emph{{Observational signatures of
  Schwarzschild-MOG black holes in scalar-tensor-vector gravity: shadows and
  rings with different accretions}},
  \href{http://dx.doi.org/10.1140/epjc/s10052-022-10868-y}{\emph{Eur. Phys. J.
  C} {\bf 82}, (2022) 885}.

\bibitem{Guo:2021bwr}
S.~Guo, K.-J. He, G.-R. Li and G.-P. Li, \emph{{The shadow and photon sphere of
  the charged black hole in Rastall gravity}},
  \href{http://dx.doi.org/10.1088/1361-6382/ac12e4}{\emph{Class. Quant. Grav.}
  {\bf 38}, (2021) 165013}, \href{http://arxiv.org/abs/2205.07242}{{\tt
  arXiv:2205.07242}}.

\bibitem{Wen:2022hkv}
S.~Wen, W.~Hong and J.~Tao, \emph{{Observational Appearances of Magnetically
  Charged Black Holes in Born-Infeld Electrodynamics}},
  \href{http://dx.doi.org/10.1140/epjc/s10052-023-11431-z}{\emph{Eur. Phys. J.
  C} {\bf 83}, (2023) 277}, \href{http://arxiv.org/abs/2212.03021}{{\tt
  arXiv:2212.03021}}.

\bibitem{Chakhchi:2022fls}
L.~Chakhchi, H.~El~Moumni and K.~Masmar, \emph{{Shadows and optical appearance
  of a power-Yang-Mills black hole surrounded by different accretion disk
  profiles}}, \href{http://dx.doi.org/10.1103/PhysRevD.105.064031}{\emph{Phys.
  Rev. D} {\bf 105}, (2022) 064031}.

\bibitem{Hou:2022eev}
Y.~Hou, Z.~Zhang, H.~Yan, M.~Guo and B.~Chen, \emph{{Image of a Kerr-Melvin
  black hole with a thin accretion disk}},
  \href{http://dx.doi.org/10.1103/PhysRevD.106.064058}{\emph{Phys. Rev. D} {\bf
  106}, (2022) 064058}, \href{http://arxiv.org/abs/2206.13744}{{\tt
  arXiv:2206.13744}}.

\bibitem{Kuang:2022xjp}
X.-M. Kuang and A.~\"Ovg\"un, \emph{{Strong gravitational lensing and shadow
  constraint from M87* of slowly rotating Kerr-like black hole}},
  \href{http://dx.doi.org/10.1016/j.aop.2022.169147}{\emph{Annals Phys.} {\bf
  447}, (2022) 169147}, \href{http://arxiv.org/abs/2205.11003}{{\tt
  arXiv:2205.11003}}.

\bibitem{Uniyal:2022vdu}
A.~Uniyal, R.~C. Pantig and A.~\"Ovg\"un, \emph{{Probing a non-linear
  electrodynamics black hole with thin accretion disk, shadow, and deflection
  angle with M87* and Sgr A* from EHT}},
  \href{http://dx.doi.org/10.1016/j.dark.2023.101178}{\emph{Phys. Dark Univ.}
  {\bf 40}, (2023) 101178}, \href{http://arxiv.org/abs/2205.11072}{{\tt
  arXiv:2205.11072}}.

\bibitem{Uniyal:2023inx}
A.~Uniyal, S.~Chakrabarti, R.~C. Pantig and A.~\"Ovg\"un, \emph{{Nonlinearly
  charged black holes: Shadow and Thin-accretion disk}},
  \href{http://arxiv.org/abs/2303.07174}{{\tt arXiv:2303.07174}}.

\bibitem{Wang:2023vcv}
X.-J. Wang, X.-M. Kuang, Y.~Meng, B.~Wang and J.-P. Wu, \emph{{Rings and images
  of Horndeski hairy black hole illuminated by various thin accretions}},
  \href{http://dx.doi.org/10.1103/PhysRevD.107.124052}{\emph{Phys. Rev. D} {\bf
  107}, (2023) 124052}, \href{http://arxiv.org/abs/2304.10015}{{\tt
  arXiv:2304.10015}}.

\bibitem{Wielgus:2021peu}
M.~Wielgus, \emph{{Photon rings of spherically symmetric black holes and robust
  tests of non-Kerr metrics}},
  \href{http://dx.doi.org/10.1103/PhysRevD.104.124058}{\emph{Phys. Rev. D} {\bf
  104}, (2021) 124058}, \href{http://arxiv.org/abs/2109.10840}{{\tt
  arXiv:2109.10840}}.

\bibitem{Bisnovatyi-Kogan:2022ujt}
G.~S. Bisnovatyi-Kogan and O.~Y. Tsupko, \emph{{Analytical study of
  higher-order ring images of the accretion disk around a black hole}},
  \href{http://dx.doi.org/10.1103/PhysRevD.105.064040}{\emph{Phys. Rev. D} {\bf
  105}, (2022) 064040}, \href{http://arxiv.org/abs/2201.01716}{{\tt
  arXiv:2201.01716}}.

\bibitem{Page:1974he}
D.~N. Page and K.~S. Thorne, \emph{{Disk-Accretion onto a Black Hole.
  Time-Averaged Structure of Accretion Disk}},
  \href{http://dx.doi.org/10.1086/152990}{\emph{Astrophys. J.} {\bf 191},
  (1974) 499--506}.

\bibitem{Guerrero:2021ues}
M.~Guerrero, G.~J. Olmo, D.~Rubiera-Garcia and D.~S.-C. G\'omez, \emph{{Shadows
  and optical appearance of black bounces illuminated by a thin accretion
  disk}}, \href{http://dx.doi.org/10.1088/1475-7516/2021/08/036}{\emph{JCAP}
  {\bf 08}, (2021) 036}, \href{http://arxiv.org/abs/2105.15073}{{\tt
  arXiv:2105.15073}}.

\bibitem{Wang:2022yvi}
H.-M. Wang, Z.-C. Lin and S.-W. Wei, \emph{{Optical appearance of
  Einstein-\AE{}ther black hole surrounded by thin disk}},
  \href{http://dx.doi.org/10.1016/j.nuclphysb.2022.116026}{\emph{Nucl. Phys. B}
  {\bf 985}, (2022) 116026}, \href{http://arxiv.org/abs/2205.13174}{{\tt
  arXiv:2205.13174}}.

\bibitem{Yang:2022btw}
J.~Yang, C.~Zhang and Y.~Ma, \emph{{Loop quantum black hole in a gravitational
  collapse model}},  \href{http://arxiv.org/abs/2211.04263}{{\tt
  arXiv:2211.04263}}.

\end{thebibliography}
%

\end{document}